\documentclass[onecolumn,preprintnumbers,amsmath,amssymb]{revtex4}
\usepackage{amsmath}
\usepackage{extarrows}
\usepackage[ruled]{algorithm2e}
\usepackage{graphicx}
\usepackage{dcolumn}
\usepackage{bm}
\usepackage[all]{xy}
\usepackage{indentfirst}
\usepackage{amsmath}
\usepackage{multirow}
\usepackage{mathrsfs}
\usepackage{bbm}
\usepackage{euscript}
\usepackage{amssymb}
\usepackage{extarrows}
\usepackage{bbm}
\usepackage[ruled]{algorithm2e}
\usepackage{graphicx}
\usepackage{dcolumn}
\usepackage{bm}
\usepackage[all]{xy}
\usepackage{indentfirst}
\usepackage{amsmath}
\usepackage{multirow}
\usepackage{mathrsfs}
\usepackage{euscript}
\usepackage{amssymb}
\begin{document}
\title{New Genuinely Multipartite Entanglement}

\author{Ming-Xing Luo
\footnote{Corresponding author: M. X. Luo (mxluo@swjtu.edu.cn)}}

\affiliation{\small{}The School of Information Science and Technology, Southwest Jiaotong University, Chengdu 610031, China
}

\begin{abstract}
The quantum entanglement as one of important resources has been verified by using different local models. There is no efficient method to verify single multipartite entanglement that is not generated by multisource quantum networks with local operations and shared randomness. Our goal in this work is to solve this problem. We firstly propose a new local model for describing all the states that can be generated by using distributed entangled states and shared randomness without classical communication.  This model is stronger than the biseparable model and implies new genuinely multipartite entanglement. With the present local model we prove that all the permutationally symmetric entangled pure states are new genuinely multipartite entanglement. We further prove that the new feature holds for all the multipartite entangled pure states in the biseparable model with the dimensions of local systems being no larger than $3$. The new multipartite entanglement is also robust against general noises. Finally, we provides a simple Bell inequality to verify new genuinely multipartite entangled pure qubit states in the present model. Our results show new insight into featuring the genuinely multipartite entanglement in the distributive scenarios.
\end{abstract}
\maketitle

\section{Introduction}

It is very important to explore distinctive features of entangled systems. For the simplest scenarios of two-particle system, Bell proposed a novel approach for explaining the paradox of Einstein-Podolsky-Rosen (EPR) \cite{EPR,Bell}. Specifically, Bell proved that bipartite quantum correlations generated by local measurements on a two-spin entanglement cannot be reproduced in any physics that satisfy the locality and casualty assumptions in the local hidden variable (LHV) theory. This kind of nonlocality is generic for bipartite entangled systems \cite{CHSH,Gis}. Different from the bipartite nonlocality, there are various kinds of multipartite nonlocalities for specific multi-particle systems [5-12]. So far, these entangled states have inspired widespread applications [13-19].

Single multipartite entanglement has experimental limits in transmission and storage because of its decoherence time \cite{CMB}. One solution is to use distributed settings, i.e., quantum networks \cite{Kim,LJL}. As a typical feature, two independent parties in a chain-shaped quantum network consisting of two entangled systems can build a new entanglement by using local operations and classical communication (LOCC) \cite{ZZH,SBP}. This kind of entanglement swapping provides an efficient method to build entangled systems in long-distant applications. Interestingly, it also implies a new non-multilocality going beyond the standard nonlocality of single entanglement [5-12]. Special nonlinear Bell-type inequalities are constructed for verifying these entangled systems such as chain-shaped or star-shaped networks \cite{BGP,RBB,Chav}, or general networks \cite{Luo}. Another method is from game theory \cite{Luo2}. Different from the standard Bell experiment \cite{Bell}, local joint measurements are allowed for each party who shares some particles involved in different entangled systems. This kind of local measurements generates multipartite quantum correlations going beyond those derived from single entangled systems. Other nonlocalities hold for cyclic quantum networks by using specific Bell theory without input assumptions \cite{Fri,Gisin2,RBBB}.

\begin{figure}
\begin{center}
\resizebox{240pt}{170pt}{\includegraphics{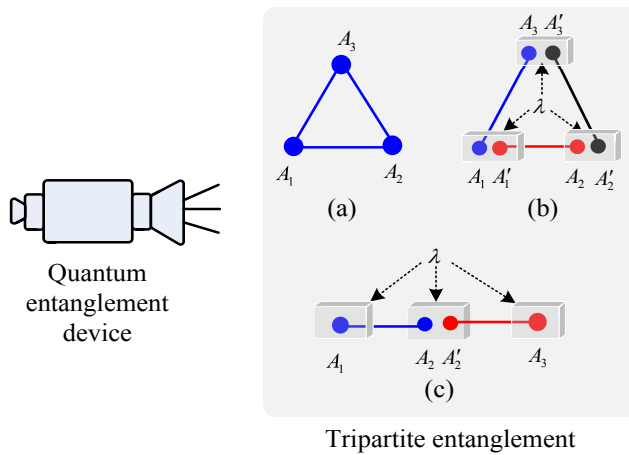}}
\end{center}
\caption{\small (Color online) Tripartite entangled systems. (a) GHZ state \cite{GHZ}. (b) Tripartite entangled system consisting of three EPR states \cite{EPR} of $|\phi\rangle_{A_1A_3},
|\phi\rangle_{A_1'A_2}$ and $|\phi\rangle_{A_2'A_3'}$. It can be regarded as a high-dimensional entangled state of $|\Phi\rangle=\frac{1}{2\sqrt{2}}(|000\rangle+|012\rangle+|120\rangle+|132\rangle+|201\rangle+|213\rangle+|321\rangle+|333\rangle)_{A_1A_1',A_2A_2',A_3A_3'}$ on Hilbert space $\mathbb{C}^4\otimes \mathbb{C}^4\otimes \mathbb{C}^4$. (c) Tripartite entangled system consisting of two EPR states $|\phi\rangle_{A_1A_2}$ and $|\phi\rangle_{A_2'A_3}$. It is equivalent to an unbalanced high-dimensional entanglement of $|\Psi\rangle=\frac{1}{2}(|000\rangle+|011\rangle+|120\rangle+|131\rangle)_{A_1,A_2A_2',A_3}$ on Hilbert space $\mathbb{C}^2\otimes \mathbb{C}^4\otimes \mathbb{C}^2$. The schematic encoding for local two-qubit system is defined according to an isomorphism mapping from Hilbert space $\mathbb{H}_{A_i}\otimes\mathbb{H}_{A_i'}$ to $\mathbb{C}^4$ as: ${\cal F}:|00\rangle\mapsto |0\rangle, |01\rangle\mapsto |1\rangle, |10\rangle\mapsto |2\rangle, |11\rangle\mapsto |3\rangle$, where $\mathbb{H}_{A_i}$ and $\mathbb{H}_{A_i'}$ are the respective state space of $A_i$ and $A_i'$. All parties are allowed to perform local operations which may depend on any shared randomness $\lambda$ in a LHV model.}
\label{fig-1}
\end{figure}

Despite of these improvements on single entangled systems and quantum networks, there is no result to distinguish single entangled systems from these being constructed by quantum networks and shared randomness. Note that classical communication is an important resource in the entanglement swapping \cite{ZZH}. However, it is not allowed for communicating local measurements with each other during Bell experiments except for the final statistics  \cite{EPR,Bell,BGP,RBB,Chav,Luo}. Otherwise, the nonlocal correlations can be forged by using shared randomness and classical communication \cite{TB}. This difference implies distinctive features for various entangled systems. One intuitive example is shown in Fig.1. All the quantum states are genuinely tripartite entangled in the biseparable model \cite{Sy} by using recent method \cite{ZDBS}, where any two parties can generate an EPR state assisted by LOCC of other parties. Surprisingly, they are inequivalent under local operations without classical communication. Actually, the Greenberger-Horne-Zeilinger (GHZ) state $|GHZ\rangle=\frac{1}{\sqrt{2}}(|000\rangle+|111\rangle)$ shown in Fig.1(a) is permutationally symmetric, i.e., the density operator is invariant under any permutation of three particles. The total system of the tripartite cyclic network shown in Fig.1(b) is equivalent to an $4$-dimensional entanglement of $|\Phi\rangle=\frac{1}{2\sqrt{2}}(|000\rangle
+|012\rangle+|120\rangle+|132\rangle
+|201\rangle+|213\rangle+|321\rangle
+|333\rangle)_{A_1A_1',A_2A_2',A_3A_3'}$ under local operations, which is invariant under the cyclic permutation of joint systems $A_1A_1', A_2A_2'$ and $A_3A_3'$ \cite{Gisin2,RBBB}. However, the total system of the tripartite chain-shaped network shown in Fig.1(c) is equivalent to the entanglement of  $|\Psi\rangle=\frac{1}{2}(|000\rangle+|011\rangle
+|120\rangle+|131\rangle)_{A_1,A_2A_2',A_3}$ under local operations, which is only invariant under the permutation of $A_1$ and $A_3$. Both $|\Phi\rangle$ and $|\Psi\rangle$ can be regarded as special cluster states \cite{Cluster,SAS}. GHZ state in Fig.1(a) cannot be generated by $|\Phi\rangle$ in Fig.1(b) or $|\Psi\rangle$ in Fig.1(c) by using local unitary operations. These features imply different nonlocalities in different local models \cite{GHZ,BGP,RBBB}. Unfortunately, it is unknown how to characterize this difference for generally entangled states.

In this work, we propose an approach to investigate new features of single multipartite entangled systems going beyond multisource quantum networks \cite{Mayer}. The main idea is to verify that some $n$-partite entangled systems with $n\geq3$ cannot be generated by using any quantum networks consisting of at most $n-1$-partite entangled states. We firstly propose a new local model for describing any state that can be generated by local operations on some quantum network consisting of at most $n-1$-partite entangled states and shared randomness without classical communication. Note that some entangled states in the biseparable model \cite{Sy} such as examples in Fig.1(b) and (c) are not entangled in the present model. This means that the present model is stronger than the biseparable model \cite{Sy}, where all the biseparable states are not entangled in the present local model. We further prove that all the permutationally symmetric $n$-partite entangled pure states such as GHZ states \cite{GHZ}, W state \cite{Dicke1} and Dicke states \cite{Toth} are new genuinely multipartite entangled in the present local model. This shows new genuinely multipartite nonlocality going beyond its verified by using the biseparable model \cite{Sy} or network model with multiple independent sources \cite{RBBB,CASA}. Moreover, we show that similar result holds for any multipartite entangled pure states in the biseparable model \cite{Sy} when all the dimensions of local systems are no more than 3. Finally, we provide a useful method to verify noisy states. These results show distinctive features of single multipartite entangled systems going beyond previous local models \cite{Sy,GHZ,Cluster,SAS,GS}.

\section{Results}

\subsection{New local model}

In this section, we propose a method to verify single entangled systems in a new local model going beyond the biseparable model \cite{Sy}. This is also interesting in fully device-independent quantum information processing \cite{Mayer,SCAK,AGC,UV}, where entanglement devices may be provided by an adversary.

Consider an $n$-partite state $\rho$ on Hilbert space $\otimes_{i=1}^n\mathbb{H}_{A_i}$, where $\mathbb{H}_{A_i}$ has local dimension $d_i$ with $d_i\geq 2$, $i=1,\cdots, n$. We define a new local model as follows.

\begin{figure}
\begin{center}
\resizebox{90pt}{120pt}{\includegraphics{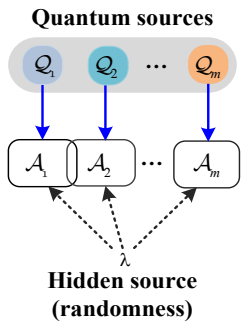}}
\end{center}
\caption{\small (Color online) Schematic local model of quantum states with distributed settings. Each quantum entangled source ${\cal Q}_i$ is shared at most $n-1$ parties ${\cal A}_i=\{\textsf{A}_{s_1}, \cdots,\textsf{A}_{s_i}\}$, $i=1, \cdots, m$. All the parties $\textsf{A}_{1}, \cdots, \textsf{A}_n$ can share a hidden randomness $\lambda$.}
\label{fig-2}
\end{figure}

\textbf{Definition 1}.  $\rho$ is new genuinely multipartite entanglement if it cannot be decomposed into a quantum network state as
\begin{eqnarray}
\rho&=&\sum_{\lambda}p(\lambda){\cal E}_1(\lambda)\otimes \cdots \otimes {\cal E}_n(\lambda)(\rho_1\otimes \cdots \otimes \rho_m)
\nonumber
\\
&=&
\sum_{\lambda}p(\lambda)\sum_{i,j}\otimes_{j=1}^nU_{i,j}(\lambda)
(\rho_1\otimes \cdots \otimes \rho_m)
(\otimes_{j=1}^nU_{i,j}(\lambda)^\dag)
\nonumber
\\
\label{eqn1}
\end{eqnarray}
where $\rho_1, \cdots, \rho_m$ are entangled states shared at most  $n-1$ parties, $\{p(\lambda)\}$ is a distribution of random variable $\lambda$, and ${\cal E}_i(\lambda)$  is a completely positive trace-preserving (CPTP mapping) \cite{GC} depending on one shared measurable variable $\lambda$ performed by the $i$-th party, which can be further represented by Kraus operators $U_{i,j}(\lambda)$ with $\sum_jU_{i,j}(\lambda)U_{i,j}(\lambda)^\dag=\mathbbm{1}$, $\mathbbm{1}$ denotes the identity operator, $i=1, \cdots, n$. Here, $U_{i,j}(\lambda)$ are quantum operations that are physically realizable without classical communication.

The classical communication is not allowed in the present local model, i.e., all the local parties cannot communicate their local operations during the process of state generation. The reason is as follows. For any two entangled pure states $|\Phi_0\rangle$ and $|\Phi_1\rangle$ on the same Hilbert space $\mathbb{H}$, we can prove that they are equivalent under LOCC assisted by teleportation-based quantum computation \cite{GC}, even if $|\Phi_0\rangle$ and $|\Phi_1\rangle$ have different nonlocalities in different local models \cite{GHZ,BGP,RBBB}.

Take the state shown in Fig.1(b) as an example. The total system has the decomposition in Eq.(\ref{eqn1}) with three bipartite entangled pure states. Moreover, the total system in Fig.1(c)  has the decomposition in Eq.(\ref{eqn1}) with two bipartite entangled pure states. Interestingly, GHZ state shown in Fig.1(a) cannot be generated according to the distributed settings in Fig.1(b) or Fig.1(c) without classical communication. It will be formally proved in Theorem 1.

Consider a general state $\rho$ on Hilbert space $\otimes_{i=1}^n\mathbb{H}_{A_i}$. If $\rho$ is a genuinely $n$-partite entanglement in the biseperable model \cite{Sy}, it cannot be decomposed into the biseparable state as
\begin{eqnarray}
\rho=\sum_{I_1, I_2}\sum_ip_i\rho^{(I_1)}_i\otimes{}\rho^{(I_2)}_i
\label{eq4}
\end{eqnarray}
where $\{I_1, I_2\}$ is a bipartite partition of $\{A_1, \cdots, A_n\}$, i.e., $I_1\cup{} I_2=\{A_1, \cdots, A_n\}$ and $I_1\cap I_2=\varnothing$, and $\rho^{(I_j)}_i$ are density operators of local system $I_j$, $j=1,2$. From Eq.(\ref{eq4}), it is easy to prove that the total states shown in Fig.1(b) and (c) are genuinely tripartite entangled states in the biseperable model \cite{Sy}. However, they are not new genuinely multipartite entangled in the local model given in Eq.(\ref{eqn1}). This means that the present multipartite entanglement from Definition 1 is stronger than its verified by using the biseperable model \cite{Sy} given in Eq.(\ref{eq4}).

Generally, consider an $n$-partite quantum network as shown in Fig.2. Each entangled source ${\cal Q}_j$ is shared by $\ell_i$ parties ${\cal A}_i=\{\textsf{A}_{s_1},  \cdots,\textsf{A}_{s_{\ell_i}}\}$ with $1\leq \ell_i<n$.  The present local model in Eq.(\ref{eqn1}) allows local quantum operations depending on any measurable variable $\lambda$  \cite{Bu}. This is different from the non-multilocality \cite{BGP,RBB,Chav,Luo} and genuinely tripartite nonlocality \cite{Fri,Gisin2,RBBB} with the assumption of multiple independent randomness shared by all parties. Unfortunately, previous Bell-type inequalities \cite{BCM,BCPS} or witness operators \cite{HHH} are difficult to verify the new genuinely multipartite entangled systems.

\subsection{New genuinely multipartite entanglement}

Our goal in this section is to explore new genuinely entangled states in the local model given in Eq.(\ref{eqn1}). Similar to the examples in Fig.1, we will prove that the lack of symmetry is actually generic for all the states with decompositions in Eq.(\ref{eqn1}). An $n$-partite state is permutationally symmetric if its density operator $\rho$ is invariant under any permutation operation $\mathsf{g}\in \mathbb{S}_n$, where $\mathbb{S}_n$ denotes the permutation group associated with a set of $n$ different elements.

One example is generalized $d$-dimensional GHZ state \cite{GHZ,Cere} defined by
\begin{eqnarray}
|GHZ\rangle=\sum_{i=0}^{d-1}a_i|i\cdots{} i\rangle_{A_1\cdots{}A_n}
\label{eqn2}
\end{eqnarray}
on Hilbert space $\otimes_{i=1}^n\mathbb{H}_{A_i}$, where $\mathbb{H}_{A_i}$s have the same local dimension $d$ with $d\geq 2$, and $a_i$s satisfy $\sum_{i=0}^{d-1}a_i^2=1$.

Another example is generalized Dicke state \cite{Toth} (including W state as a special example \cite{Dicke1}) given by
\begin{eqnarray}
|D_{k,n}\rangle=\frac{1}{\sqrt{N_{k,n}}}\sum_{i_1+\cdots{}
+i_n=k}|i_1 \cdots{}i_n\rangle_{A_1\cdots{}A_n}
\label{eqn3}
\end{eqnarray}
where $N_{k,n}=C(k+n-1,n-1)$ denotes the combination number of choosing $n-1$ balls from a box with $k+n-1$ balls, $k=1, \cdots, nd-d-1$. Dicke states are very interesting because of the robustness against the particle loss, global dephasing, and bit flip noise \cite{Dicke2}. These states are genuinely multipartite entangled \cite{DPR,Dicke4,TDS} in the biseparable model \cite{Sy} given in Eq.(\ref{eq4}). They are also different from cluster states \cite{Cluster} or stabilize states \cite{HEB}. Generally, any permutationally symmetric pure state can be represented by the superposition of Dicke states. All the permutationally symmetric entangled pure states can be verified using Hardy-type inequalities \cite{CYZ} in the biseparable model \cite{Sy}. Here, we show that they are new genuinely entangled states in the present model in Definition 1.

{\bf Theorem 1}. Any $n$-partite ($n\geq 3$) permutationally symmetric entangled pure state in the biseparable model is new genuinely $n$-partite entanglement in the local model given in Eq.(\ref{eqn1}), i.e., it cannot be generated by using $m$-partite entangled states with $m<n$ under local operations and shared randomness.

The product state $|\phi\rangle^{\otimes n}$, which is permutationally symmetric, is excluded in Theorem 1.

To show the main idea of proof, we take $4$-dimensional  permutationally symmetric entangled state $|G\rangle=\frac{1}{2}\sum_{i=0}^3|iii\rangle_{A_1B_1C_1}$ as an example. Here, all the $4$-dimensional basis states of $A_1$ ($B_1$, or $C_1$) can be represented by two qubits $A_{11}$ and $A_{12}$ ($B_{11}$ and $B_{12}$, or $C_{11}$ and $C_{12}$) according to the inverse mapping of ${\cal F}$ defined in Fig.1, where $\mathbb{H}_{A_1}=\mathbb{H}_{A_{11}}\otimes \mathbb{H}_{A_{12}}$, $\mathbb{H}_{B_1}=\mathbb{H}_{B_{11}}\otimes \mathbb{H}_{B_{12}}$ and $\mathbb{H}_{C_1}=\mathbb{H}_{C_{11}}\otimes \mathbb{H}_{C_{12}}$. Similar to Eq.(\ref{eqn1}), if $|G\rangle$ can be decomposed into two states under the local operations $W_i$, we can prove that $|G\rangle$ is actually decomposed into two tripartite GHZ states $|G_1\rangle=\frac{1}{\sqrt{2}}(|000\rangle+|111\rangle)$ and $|G_2\rangle=\frac{1}{\sqrt{2}}(|000\rangle+|111\rangle)$, i.e., $(W_1\otimes {}W_2\otimes{}W_3) |G\rangle=|G_1\rangle_{A_{11}B_{11}C_{11}}|G_2\rangle_{A_{12}B_{12}C_{12}}$. Moreover, one of $|G_1\rangle$ and $|G_2\rangle$ ($|G_1\rangle$ for example) is also tripartite entangled in the biseparable model \cite{Sy}. Note that a 2-dimensional Hilbert space cannot be further decomposed into the tensor of two Hilbert spaces with at least two dimensions. This implies that $|G_1\rangle$ cannot be further decomposed into two states. So, there is a tripartite entanglement in the biseparable model \cite{Sy} after decomposing $|G\rangle$ under any local operations. It means that $|G\rangle$ must be generated by some states $|\Psi_1\rangle, \cdots, |\Psi_m\rangle$ satisfying that one of $|\Psi_i\rangle$s is a tripartite entanglement in the biseparable model \cite{Sy}. $|G\rangle$ has no decomposition in Eq.(\ref{eqn1}) under the local operations.

Generally, for any $n$-partite ($n\geq 3$) permutationally symmetric entangled pure state $|\Phi\rangle$ in the biseparable model \cite{Sy}, we prove that there is an $n$-partite entanglement after any decomposition of $|\Phi\rangle$ under local operations. It means that $|\Phi\rangle$ must be generated by some states $|\Psi_1\rangle, \cdots, |\Psi_m\rangle$ under local operations and shared randomness, where one of $|\Psi_i\rangle$s is an $n$-partite entanglement in the biseparable model \cite{Sy}. Hence, $|\Phi\rangle$ has no decomposition in Eq.(\ref{eqn1}), where all the decomposed states are at most $n-1$-partite entanglement in the biseparable model \cite{Sy}. The detailed proof of Theorem 1 is shown in Supplementary A.

Theorem 1 implies that any multipartite permutationally symmetric entangled pure state shows a new kind of $n$-partite nonlocality. It is well-known that LOCC cannot increase the entanglement of states \cite{HHH}. Our result shows a further feature of classical communication for distinguishing single entangled systems from these generated by distributed settings without classical communication, even if these states may be equivalent under LOCC (see examples in Fig.1). The new genuinely multipartite entangled systems may have special symmetry under local unitary operations \cite{GHZ,Cere,Toth,Dicke1,Dicke2}. Interestingly, there are general systems that are also new genuinely multipartite entangled in the local model given in Eq.(\ref{eqn1}).

{\bf Theorem 2}. Any $n$-partite ($n\geq3$) entangled pure state on Hilbert space $\otimes_{i=1}^n\mathbb{H}_{A_i}$ in the biseparable model in Eq.(\ref{eq4}) is new genuinely $n$-partite entanglement in the local model given in Eq.(\ref{eqn1}) if the dimension of $\mathbb{H}_{A_i}$ is no larger than $3$, $i=1, \cdots, n$.

The proof of Theorem 2 is shown in Supplementary B. The main idea is that each Hilbert space $\mathbb{H}$ with local dimension $d\leq 3$ cannot be decomposed into the tensor of two Hilbert spaces $\mathbb{H}_1$ and $\mathbb{H}_2$ with at least two dimensions (even if under the isomorphism mappings). Theorem 2 is important for hybrid systems whose local particles have different state spaces \cite{KBK}. Theorems 1 and 2 show general results of new genuinely multipartite entanglement in the present model given in Eq.(\ref{eqn1}). A directive result is for any pure state being equivalent to one entangled state in Theorem 1 or 2 using local unitary operations and auxiliary states.

\begin{figure}
\begin{center}
\resizebox{90pt}{100pt}{\includegraphics{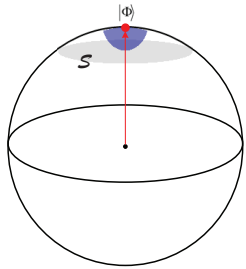}}
\end{center}
\caption{\small (Color online) Schematic entangled states on Bloch sphere. The red dot denotes the pure state $|\Phi\rangle$. $|\Phi\rangle$ is a new genuinely $n$-partite entangled in the local model given in Eq.(\ref{eqn1}). The grey hyperplane ${\cal S}$ is defined by all the network states $\rho_{net}$ defined in Eq.(\ref{eqn1}) with $F(\rho_{net},|\Phi\rangle)=D(|\Phi\rangle)$, where $D(|\Phi\rangle)$ and $F(\rho_{net},|\Phi\rangle)$ are defined in Eq.(\ref{eqn9}). All the states above the grey hyperplane are new genuinely multipartite entangled in the local model given in Eq.(\ref{eqn1}).}
\label{fig-3}
\end{figure}

\subsection{New genuinely multipartite entanglement with noises}

In this section, we will prove that the present new genuinely multipartite entanglement is robust against general noises. For an $n$-partite pure state on Hilbert space $\otimes_{i=1}^n\mathbb{H}_{A_i}$, inspired by the biseparable model \cite{Sy,HHH}, denote $D(|\Phi\rangle)$ as the maximal distance between $|\Phi\rangle$ and all the network states $\rho_{net}$ defined in Eq.(\ref{eqn1}), i.e.,
\begin{eqnarray}
D(|\Phi\rangle)=\sup_{\rho_{net}}F(\rho_{net},\rho_{\Phi})
\label{eqn9}
\end{eqnarray}
where $\rho_{\Phi}$ denotes the density operator defined by $\rho_{\Phi}=|\Phi\rangle\langle \Phi|$, $F(\rho,\rho_{\Phi})=({\rm tr}(|\Phi\rangle\langle \Phi|\sqrt{\rho}))^2$ denotes the fidelity between $\rho$ and $\rho_{\Phi}$ \cite{Jozsa}. It is easy to get that $F(\rho_{net},\rho_{\Phi})\leq \max_{i}|\langle \Phi|\Psi_i\rangle|^2$ if $\rho_{net}$ has the decomposition of $\rho_{net}=\sum_ip_i|\Psi_i\rangle\langle\Psi_i|$. Hence, it is sufficient to evaluate $ D(|\Phi\rangle)$ by considering all the pure states with the decompositions in Eq.(\ref{eqn1}) as
\begin{eqnarray}
D(|\Phi\rangle)=\sup_{\otimes_{i=1}^m|\Psi_i\rangle}|\langle \Phi|\otimes_{j=1}^nU_j (\otimes_{i=1}^m| \Psi_i\rangle)|^2
\label{eqn10}
\end{eqnarray}
where $|\Psi_i\rangle$ is any $n$-partite states with at most $n-1$ particles being entangled, $U_i$ is an arbitrary local unitary operation performed by the $i$-th party, $i=1, \cdots, n$.

From the compactness of Bloch sphere, using the present model given in Eq.(\ref{eqn1}) we get
\begin{eqnarray}
D(|\Phi\rangle)<1
\label{eqn11}
\end{eqnarray}
for any new genuinely multipartite entangled state $|\Phi\rangle$. From Fig.3, we get a sufficient condition for witnessing a new genuinely multipartite entangled state $\rho$ in the local model given in Eq.(\ref{eqn1}) if
\begin{eqnarray}
F(\rho,\rho_{\Phi})>D(|\Phi\rangle)
\label{eqn12}
\end{eqnarray}

Note that from the convexity of fully separable states or biseparable states \cite{Sy,HHH}, there is a witness operator \cite{HHH,Chru} for verifying each entanglement apart from fully separable states, or each genuinely multipartite entanglement in the biseparable model \cite{Sy}. Interestingly, all the states defined in Eq.(\ref{eqn1}) consist of a convex set. Hence, the condition in Eq.(\ref{eqn12}) implies a new witness operator \cite{HHH,Chru} as
\begin{eqnarray}
{\cal W}=F(\rho,\rho_{\Phi})\mathbb{I}-|\Phi\rangle\langle \Phi|
\label{eqnw}
\end{eqnarray}
in order to separate a multipartite state $\rho$ from all the network states defined in Eq.(\ref{eqn1}), i.e., we get from Eqs.(\ref{eqn10})-(\ref{eqn12}) that \begin{eqnarray}
{\rm tr}[{\cal W}\rho_{net}]\geq 0
\label{eqnw1}
\end{eqnarray}
for all the network states $\rho_{net}$ defined in Eq.(\ref{eqn1}), and
\begin{eqnarray}
{\rm tr}[{\cal W}\varrho]< 0
\label{eqnw2}
\end{eqnarray}

In applications, it is sufficient to choose a specific multipartite entangled pure state $|\Phi\rangle$ in the local model given, which closes to $\varrho$.

Although it is difficult to evaluate $D(|\Phi\rangle)$ for general states, we provide some sufficient conditions for entangled states in Theorems 1 and 2. Define a generalized permutationally symmetric entanglement as
\begin{eqnarray}
|\Phi_{ps}\rangle=\sum_{i=0}^{nd-n-1}
\alpha_i|D_{i,n}\rangle
\label{eqn13}
\end{eqnarray}
where $|D_{0,n}\rangle:=\beta_0|0\rangle^{\otimes n}+\beta_{1}|d-1\rangle^{\otimes n}$ with $\beta_0^2+\beta_1^2=1$, and $|D_{k,n}\rangle$ is Dicke states defined in Eq.(\ref{eqn3}). Here, special restrictions of $\alpha_i$s should be imposed to exclude symmetric product states. Generally, we prove the following result.

{\bf Theorem 3}. For any $n$-partite ($n\geq3$) state $\rho$ on Hilbert space $\otimes_{i=1}^n\mathbb{H}_{A_i}$, it is new genuinely entangled in the model given in Eq.(\ref{eqn1}) if one of the following facts holds
\begin{itemize}
\item[(i)]For generalized GHZ state $|GHZ\rangle$ defined in Eq.(\ref{eqn2}), $\rho$ satisfies
\begin{eqnarray}
F(\rho,\rho_{GHZ})>\max\{a^2_0,\cdots, a^2_{d-1}\}
\label{eqn14}
\end{eqnarray}
\item[(ii)] For Dicke state $|D_{k,n}\rangle$ defined in Eq.(\ref{eqn3}), $\rho$ satisfies
\begin{eqnarray}
F(\rho,\rho_{D_{k,n}})&=&F(\rho,\rho_{D_{nd-n-k-1,n}})
\nonumber\\
&>&\frac{n-1}{n+k-1}
\label{eqn15}
\end{eqnarray}
for $k=1, \cdots, \lfloor\frac{nd-n-1}{2}\rfloor$, where $\lfloor{}x\rfloor$ denotes the maximal integer no larger than $x$.
\item[(iii)] For $|\Phi_{ps}\rangle$ defined in Eq.(\ref{eqn13}), $\rho$ satisfies
\begin{eqnarray}
F(\rho,\rho_{\Phi_{ps}})
&>&
\sum_{i=1}^{\lfloor\frac{nd-n-1}{2}\rfloor}
\frac{n-1}{n+i-1}(\alpha_i^2+\alpha_{nd-n-1-i}^2)
\nonumber
\\
&&+\alpha_0^2\beta^2
\label{eqn16a}
\end{eqnarray}
 for an odd $nd-n-1$; or
\begin{eqnarray}
F(\rho,\rho_{\Phi_{ps}})
&>&\sum_{i=1}^{\lfloor\frac{nd-n-1}{2}\rfloor}
\frac{n-1}{n+i-1}(\alpha_i^2+\alpha_{n-i}^2)
\nonumber
\\
&&+\alpha_0^2\beta^2-\frac{2n-2}{nd+n-3}\alpha_{\frac{nd-n-1}{2}}^2
\label{eqn16b}
\end{eqnarray}
for an even $nd-n-1$, where  $\beta$ is given by $\beta=\max\{\beta_0,\beta_1\}$.
\item[(iv)] For a new genuinely $n$-partite entangled qubit state $|\Phi\rangle$ in the present model, $\rho$ satisfies
\begin{eqnarray}
F(\rho,\rho_{\Phi})>\max_{{\cal A}\subset\{A_1, \cdots, A_n\}}\max\{\sigma(\rho_{\cal A})\}
\label{eqn17}
\end{eqnarray}
 where $\sigma(\rho_{\cal A})$ denotes all the eigenvalues of $\rho_{\cal A}$, and $\rho_{\cal A}$ denotes the reduced density matrix $\rho_{\cal A}$ of the subsystems in ${\cal A}$ with ${\cal A}\subset\{A_1, \cdots, A_n\}$.
 \end{itemize}

Theorem 3 provides a useful method to verify new genuinely multipartite entangled states with general noises in the local model given in Eq.(\ref{eqn1}). The proof of Theorem 3 is given in Supplementary C. The main idea is to evaluate $D(|\Phi\rangle)$ defined in Eq.(\ref{eqn10}) for special states $|\Phi\rangle$.

\subsection{Examples}

In this section, we present some examples of new genuinely multipartite entangled states with noises.

\textbf{Example 1}. Consider an $n$-partite GHZ state defined in Eq.(\ref{eqn2}) with white noise as follows \cite{Werner}:
\begin{eqnarray}
\rho_v=v|GHZ\rangle\langle GHZ|+\frac{1-v}{d^n}\mathbbm{1}_{d^n}
\label{eqn18}
\end{eqnarray}
where $\mathbbm{1}_{d^n}$ denotes the identity operator on Hilbert space $\otimes_{i=1}^n\mathbb{H}_{A_i}$, and $\mathbb{H}_{A_i}$ has the same local dimension $d$ with $d\geq 2$. The weight $v$ may operationally represent the interferometric contrast observed in experiment \cite{Werner}. For the maximally entangled GHZ state $|GHZ\rangle$ with $d=2$, from Eq.(\ref{eqn14}), $\rho_v$ is new genuinely $n$-partite entangled in the local model given in Eq.(\ref{eqn1}) for $v>\frac{2^{n-1}-1}{2^{n}-1}$. This is consistent with the genuinely multipartite nonlocality in the biseparable model \cite{GS,Sy} given in Eq.(\ref{eq4}). Moreover, from Eqs.(\ref{eqnw})-(\ref{eqnw2}) and Eq.(\ref{eqn14}), it follows a witness operator as
\begin{eqnarray}
{\cal W}_{GHZ}=a\mathbbm{1}_{d^n}-|GHZ\rangle\langle GHZ|
\end{eqnarray}
for verifying $\rho_v$ in the local model given in Definition 1 when
\begin{eqnarray}
v\geq \frac{d^na-1}{d^n-1}
\label{eqn19}
\end{eqnarray}
where $a=\max\{a^2_0, \cdots, a_{d-1}^2\}$.

\textbf{Example 2}. Consider a Dicke state defined in Eq.(\ref{eqn3}) with white noise as follows
\begin{eqnarray}
\varrho_v=v|D_{k,n}\rangle\langle{}D_{k,n}|
+\frac{1-v}{d^n}\mathbbm{1}_{d^n}
\label{eqn20}
\end{eqnarray}
where $v\in [0,1]$, and $k\leq \lfloor\frac{nd-n-1}{2}\rfloor$. From Eqs.(\ref{eqnw})-(\ref{eqnw2}) and (\ref{eqn15}), it provides a witness operator as
\begin{eqnarray}
{\cal W}_{D_{k,n}}=\frac{n-1}{n+k-1}\mathbbm{1}_{d^n}-|D_{k,n}\rangle\langle D_{k,n}|
\label{eqn21w}
\end{eqnarray}
for verifying new genuinely $n$-partite entangled $\varrho_v$ in the local model given in Eq.(\ref{eqn1}) when $v$ satisfies
\begin{eqnarray}
v\geq \frac{(n-1)2^n-n-k+1}{(2^n-1)(n+k-1)}
\label{eqn21}
\end{eqnarray}

Similarly, we can verify a general permutationally symmetric noisy state $\varrho_v$ defined by
\begin{eqnarray}
\varrho_v=\sum_{k=1}^{nd-n-1}v_k|D_{k,n}\rangle\langle{}D_{k,n}|+\frac{v_0}{d^n}
\mathbbm{1}_{d^n}
\label{eqn22}
\end{eqnarray}
where $v_i\geq0$ and $\sum_{i=0}^{nd-n-1}v_i=1$. One method is from Eq.(\ref{eqn21w}). Define an operator ${\cal W}_{D,n}$ as
\begin{eqnarray}
{\cal W}_{D,n}=\sum_{k=1}^{nd-n-1}{\cal W}_{D_{k,n}}
\label{eqn22w}
\end{eqnarray}
It is forward to prove that ${\cal W}_{D,n}$ is a useful witness operator from Eqs.(\ref{eqnw})-(\ref{eqnw2}) when $v_0$ satisfies the following inequality
\begin{eqnarray}
v_0<\frac{d^n}{d^n-nd+n+1}(1-\sum_{k=1}^{nd-n-1}L_k)
\label{eqn22wb}
\end{eqnarray}
where $L_k=L_{nd-n-1-k}=\frac{n-1}{n+k-1}$ for $k=1$, $\cdots$, $\lfloor\frac{nd-n-1}{2}\rfloor$.

Another way for verifying $\varrho_v$ in Eq.(\ref{eqn22}) is using permutationally symmetric state $|\Phi_{ps}\rangle$ defined in Eq.(\ref{eqn13}). In fact, from Eq.(\ref{eqn16a}), it is easy to construct a witness operator ${\cal W}_{\Phi_{ps}}$ for verifying $\varrho_v$ in the local model given in Eq.(\ref{eqn1}) as
\begin{eqnarray}
{\cal W}_{\Phi_{ps}}
&=&(
\sum_{i=1}^{\lfloor\frac{nd-n-1}{2}\rfloor}
\frac{n-1}{n+i-1}(\alpha_i^2+\alpha_{nd-n-1-i}^2)
\nonumber
\\
&&
\left.+\alpha_0^2\beta^2\right)\mathbbm{1}_{d^n}-|\Phi_{ps}\rangle\Phi_{ps}|
\label{eqn22wc}
\end{eqnarray}
when all $v_i$s satisfy
\begin{eqnarray}
\sum_{k=1}^{nd-n-1}v_k\alpha_k^2+\frac{v_0\alpha_0^2}{d^n}&>&
\sum_{i=1}^{\lfloor\frac{nd-n-1}{2}\rfloor}
\frac{n-1}{n+i-1}(\alpha_i^2+\alpha_{nd-n-1-i}^2)
\nonumber
\\
&&
+\alpha_0^2\beta^2
\label{eqn22wd}
\end{eqnarray}
Similar result holds for $nd-n-1$ being an even integer from Eq.(\ref{eqn16b}).

Note that the present model given in Eq.(\ref{eqn1}) is stronger than the biseperable model in Eq.(\ref{eq4}). All the witness operators of ${\cal W}_{D_{k,n}}$ defined in Eq.(\ref{eqn21w}), ${\cal W}_{D,n}$ defined in Eq.(\ref{eqn22w}) and ${\cal W}_{\Phi_{ps}}$ defined in Eq.(\ref{eqn22wc}) are then useful for verifying genuinely multipartite entangled states in the biseparable model \cite{Sy}.

\textbf{Example 3}. Consider a new genuinely $n$-partite entangled pure state $|\Phi\rangle$ on Hilbert space $\otimes_{i=1}^n\mathbb{H}_{A_i}$ in the local model given in Eq.(\ref{eqn1}), where $d_i$ denotes the dimension of $\mathbb{H}_{A_i}$, $i=1, \cdots, n$. Define a noisy state of $|\Phi\rangle$ as
\begin{eqnarray}
\rho_v=v|\Phi\rangle\langle \Phi|+(1-v)\varrho
\label{eqn23}
\end{eqnarray}
where $\varrho$ is a general noisy operator, such as white noise \cite{Werner}, depolarization \cite{BGK}, or erasure errors \cite{GBP}, which should be positive semidefinite with unit trace. Theorem 3 provides an efficient method to verify $\rho_v$ in Eq.(\ref{eqn23}) in the local model given in Eq.(\ref{eqn1}). One method is to use the symmetric space spanned by Dicke states $|D_{k,n}\rangle$. Especially, by using the witness operator ${\cal W}_{D,n}$ defined in Eq.(\ref{eqn22w}), there is a sufficient condition
\begin{eqnarray}
v>\frac{\sum_{k=1}^{nd-n-1}L_k}{\sum_{k=1}^{nd-n-1}\beta_k}
\label{eqn24}
\end{eqnarray}
in order to verify a new genuinely $n$-partite entangled $\rho_v$ in Eq.(\ref{eqn23}) in the local model given in Eq.(\ref{eqn1}), where $\beta_i=|\langle{}D_{i,n}|\Phi\rangle|^2$ for $i=1, \cdots, nd-n-1$, and $L_k=L_{nd-n-1-k}=\frac{n-1}{n+k-1}$ for $k=1, \cdots, \lfloor\frac{nd-n-1}{2}\rfloor$.

The second is using the witness operator ${\cal W}_{\Phi_{ps}}$ defined in Eq.(\ref{eqn22wc}) when $v$ satisfies
\begin{eqnarray}
v>\alpha_0^2\beta^2+
\sum_{i=1}^{\lfloor\frac{nd-n-1}{2}\rfloor}
\frac{n-1}{n+i-1}(\alpha_i^2+\alpha_{nd-n-1-i}^2)
\label{eqn24}
\end{eqnarray}
These conditions are also useful for verifying the genuinely multipartite entanglement in the biseparable model \cite{Sy}.

Another one is using the condition defined in Eq.(\ref{eqn17}) qubit states, $i=1, \cdots, 3$. Assume that $|\Phi\rangle$ is given by
\begin{eqnarray}
|\Phi\rangle=\sum_{i_1\cdots i_n}\alpha_{i_1\cdots i_n}|i_1\cdots{}i_n\rangle
\label{F1}
\end{eqnarray}
For the bipartition $\{A_1, \cdots, A_k\}$ and $\{A_{k+1}, \cdots, A_n\}$ of $\{A_1, \cdots, A_n\}$, denote $\rho_{A_1\cdots{}A_k}$ as the reduced density matrix on the subsystems $\{A_1, \cdots, A_k\}$. It follows that \cite{HJ}:
\begin{eqnarray}
\lambda_{\max}(\rho_{A_1\cdots{}A_k})
\leq \max\{\|v_i\|_1\}
\label{F3}
\end{eqnarray}
where $v_j$ is the $j$-th column vector of $\rho_{A_1\cdots{}A_k}$, and $\|x\|=\sum_{i}|x_i|$ denotes the $1$-norm of vector. This provides an efficient method without evaluating all the eigenvalues of the reduced density matrices. From Eqs.(\ref{eqn17}) and (\ref{eqn25c}) we obtain a simple witness operator as
\begin{eqnarray}
{\cal W}_{\Phi}=\alpha \mathbbm{1}_{d_1\cdots{}d_n}-|\Phi\rangle\langle \Phi|
\label{eqn27a}
\end{eqnarray}
for verifying new genuinely $n$-partite entanglement in the local model given in Eq.(\ref{eqn1}), where $\alpha=\max\{\|\rho_{{\cal A}}\|_c\}$, and $\|\rho_{{\cal A}}\|_c$ denotes the maximal $1$-norm of column vectors in the reduced density matrix $\rho_{{\cal A}}$ with ${\cal A}\subset \{A_1, \cdots, A_n\}$, and  $\mathbbm{1}_{d_1\cdots{}d_n}$ denotes the identity operator on $\otimes_{i=1}^n\mathbb{H}_{A_i}$. The witness operator ${\cal W}_{\Phi}$ implies a sufficient condition as
\begin{eqnarray}
v>\frac{\alpha-{\rm tr}[\varrho|\Phi\rangle\langle \Phi|]}{1-{\rm tr}[\varrho|\Phi\rangle\langle \Phi|]}
\label{eqn30}
\end{eqnarray}
for verifying $\rho_v$ being a new genuinely $n$-partite entanglement in the present model given in Eq.(\ref{eqn1}).

\textbf{Example 4}. Consider a three-qubit entangled pure state as \cite{AAC,Sy}:
\begin{eqnarray}
|\Phi_3\rangle_{ABC}
&=&\lambda_0|000\rangle+\lambda_1e^{i\phi}|100\rangle+\lambda_2|101\rangle
\nonumber\\
&&+\lambda_3|110\rangle
+\lambda_4|111\rangle
\label{eqn25}
\end{eqnarray}
where $\phi\in[0, \pi]$, $\lambda_i\geq 0$, and $\sum_{i=0}^4\lambda_i^2=1$. From Theorem 2, $|\Phi_3\rangle_{ABC}$ is a new genuinely $n$-partite entanglement in the present model given in Eq.(\ref{eqn1}). With some evaluations (Supplementary D), from Eq.(\ref{eqn17}) we get that any state $\rho$ is a new genuinely multipartite entanglement in the model given in Eq.(\ref{eqn1}) if
\begin{eqnarray}
F(\rho,\rho_{\Phi_3})> \gamma
\label{eqn25a}
\end{eqnarray}
where $\gamma=\max\{\gamma_1, \gamma_2, \gamma_3\}$ and $\gamma_i$ are given by $\gamma_1=\frac{1}{2}+\frac{1}{2}(1-4\lambda_0^2\lambda_2^2-4\lambda_0^2\lambda_3^2
-4\lambda_0^2\lambda_4^2)^{1/2}$, $\gamma_2=\frac{1}{2}+\frac{1}{2}
(1+4\delta-4\lambda_0^2\lambda_3^2)^{1/2}$, $\gamma_3=\frac{1}{2}+\frac{1}{2}
(1+4\delta-4\lambda_0^2\lambda_2^2)^{1/2}$ with $\delta=2\lambda_1\lambda_2\lambda_3\lambda_4-\lambda_0^2\lambda_4^2-\lambda_1^2\lambda_4^2-\lambda_2^2\lambda_3^2$.

Now, consider a noisy state as
\begin{eqnarray}
\rho_v=v|\Phi_3\rangle\langle \Phi_3|+(1-v)\varrho
\label{eqn25b}
\end{eqnarray}
where $|\Phi_3\rangle$ is defined in Eq.(\ref{eqn25}), $\varrho$ is a general density operator, and $v\in [0,1]$. This state is a special example of the state defined in Eq.(\ref{eqn23}). From Eqs.(\ref{eqnw}) and (\ref{eqn25a}) we get a witness operator as
\begin{eqnarray}
{\cal W}_{\Phi_3}=\gamma \mathbbm{1}_8-|\Phi_3\rangle\langle \Phi_3|
\label{eqn25c}
\end{eqnarray}
for verifying $\rho_v$ defined in Eq.(\ref{eqn25b}) in the local model given in Eq.(\ref{eqn1}). It provides a sufficient condition as
\begin{eqnarray}
v>\frac{\gamma-{\rm tr}[\varrho|\Phi_3\rangle\langle \Phi_3|]}{1-{\rm tr}[\varrho|\Phi_3\rangle\langle \Phi_3|]}
\label{eqn26}
\end{eqnarray}
for witnessing $\rho_v$ defined in Eq.(\ref{eqn25b}) in the present model given in Eq.(\ref{eqn1}). Similar result holds for other entangled states from Eq.(\ref{eqn25a}). Of course, the computation complexity of this method depends on the number of involved quantum systems.

\begin{figure}
\begin{center}
\resizebox{240pt}{120pt}{\includegraphics{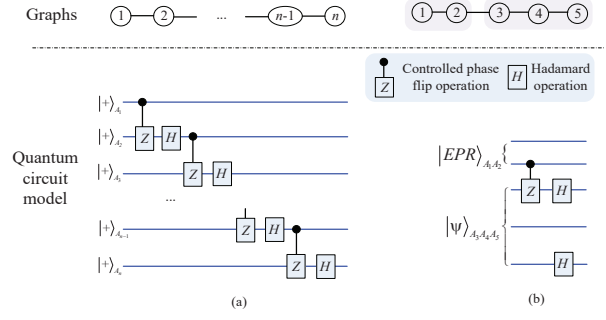}}
\end{center}
\caption{\small (Color online) Schematic cluster states. (a) $n$-qubit GHZ state. (b) Five-qubit linear cluster state. $H$ denotes Hadamard operation: $|0\rangle\mapsto |+\rangle$, and $|1\rangle\mapsto |-\rangle$ with $|\pm\rangle=\frac{1}{\sqrt{2}}(|0\rangle+|1\rangle)$. The controlled phase flip operation is given by: $|0\rangle\langle 0|\otimes \mathbbm{1}+|1\rangle\langle 1|\otimes \sigma_z$, where $\sigma_z$ is Pauli operation. }
\label{fig-4}
\end{figure}

\textbf{Example 5}. Consider cluster states as shown in Fig.\ref{fig-4}. In Fig.\ref{fig-4}(a), the final state after all the controlled phase gates is given by
\begin{eqnarray}
|\Phi\rangle=\frac{1}{\sqrt{2}}
(|0\cdots{}0\rangle+|1\cdots{}1\rangle)_{A_1\cdots{}A_n}
\label{F4}
\end{eqnarray}
which is an $n$-partite GHZ state \cite{GHZ}. The inequality (\ref{eqn19}) provides a sufficient condition to verify $|\Phi\rangle$ with white noise being a new genuinely $n$-partite entanglement in the local model given in Eq.(\ref{eqn1}).

In Fig.\ref{fig-4}(b), the final state after all controlled phase gates performed on the joint system of
$|EPR\rangle_{A_1A_2}|\Psi\rangle_{A_3A_4A_5}$ is given by
\begin{eqnarray}
|\Phi_5\rangle=\frac{1}{\sqrt{2}}
(a|00000\rangle+b|11100\rangle+a|00111\rangle
+b|11011\rangle)
\label{F7}
\end{eqnarray}
where $|\Psi\rangle_{A_3A_4A_5}$ is generalized three-qubit linear cluster state given by $|\Psi\rangle_{A_3A_4A_5}=\frac{1}{\sqrt{2}}(|+\rangle|0\rangle|+\rangle+|-\rangle|1\rangle|-\rangle)$, and $|EPR\rangle_{A_1A_2}$ denotes the generalized EPR state given by $|EPR\rangle_{A_1A_2}=a|00\rangle+b|11\rangle$ with $a^2+b^2=1$. From Theorem 2, this state is a new genuinely $5$-partite entangled state in the local model given in Eq.(\ref{eqn1}). Consider the noisy state of $\rho_{v}$ defined in Eq.(\ref{eqn23}) with $|\Phi\rangle=|\Phi_5\rangle$ and white noise $\varrho=\frac{1}{32}\mathbbm{1}_{32}$. From the inequality (\ref{eqn30}) we get a sufficient condition of $v\geq \frac{16a^2-1}{31}$ ($a\geq\frac{1}{\sqrt{2}}$) for witnessing  $\rho_{v}$ in the present model given in Eq.(\ref{eqn1}).

Generally, the inequality (\ref{eqn30}) is useful for verifying the cluster states or graph states \cite{Cluster} associated with specific graphs. These states are interesting in quantum information processing and measurement-based quantum computations \cite{Cluster,NMDB}. It is known that single connected graph states have the genuinely multipartite nonlocalities in the biseparable model \cite{GTHB,Sy}. Theorems 2 and 3 imply new generic nonlocality for the connected graph states with qubit states as inputs in the present model given in Eq.(\ref{eqn1}).

\begin{figure}
\begin{center}
\resizebox{160pt}{180pt}{\includegraphics{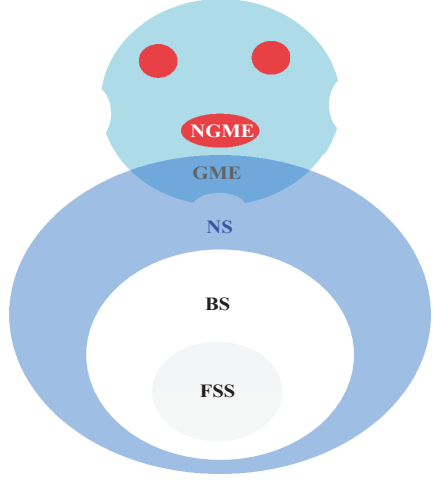}}
\end{center}
\caption{\small (Color online) Schematic relations of different states. The new genuinely multipartite entanglement (NGME) is stronger than previous genuinely multipartite entanglement (GME) verified by using the biseparable model \cite{Sy}. NS denotes network state defined in Eq.(\ref{eqn1}). BS denotes states biseparable state in the biseparable model \cite{Sy}. FSS denotes fully separable state. NGMEs (in red area) are genuinely multipartite entangled states in the biseparable model \cite{Sy}. The converse is not right, see Fig.1(b) and (c) in the main text. BSs are network states defined in Eq.(\ref{eqn1}). The converse is not right, see Fig.1(b) and (c).}
\label{fig-5}
\end{figure}

\section*{Discussion}

The present model given in Eq.(\ref{eqn1}) is derived from the distributed settings of entanglement generations \cite{ZZH,Kim}. All the entangled states that have the decompositions in Eq.(\ref{eqn1}) are not symmetric from Theorem 1. This geometric feature has a strong restriction on the resultant from a generation procedure based on quantum networks \cite{TDS} without classical communication. Interestingly, this shows a new kind of genuinely multipartite entanglement going beyond its verified by all the previous local models \cite{Bell,GHZ,Sy}. Specially, the present model in Eq.(\ref{eqn1}) is stronger than the previous biseparable model \cite{Sy} that is used to verify the genuinely multipartite entanglement, as shown in Fig.\ref{fig-5}. It means that all the new genuinely multipartite entangled states in the present model in Definition 1 are genuinely multipartite entangled states in the biseparable model \cite{Sy}. The converse is not right, see examples in Fig.1(b) and (c). Theorem 1 presents some interesting examples includes all the generalized GHZ states, Dicke states, and generalized permutationally symmetric entangled pure states which are new genuinely multipartite entangled states in the present model.

Similar results hold for other entangled pure states in the biseparable model \cite{Sy} with small dimensions of local systems from Theorem 2. This means that the present model in Eq.(\ref{eqn1}) can verify the same kind of entangled pure states with the biseparable model \cite{Sy}. Although there are proper isomorphism transformations by local parties for embedding a small state space into a large Hilbert space with the tensor decomposition, however, these local operations are useless for generating specific entangled systems without classical communication. This fact cannot be extended for high-dimensional systems, see examples in Fig.1(b) and (c). Generally, there are lots of multipartite high-dimensional states which have different features in the present model in Eq.(\ref{eqn1}) and the biseparable model \cite{Sy}. So, an interesting problem is how to feature these high-dimensional systems with the present local model. One possible solution is to explore the entangled pure states with prime dimensions of local systems. Another problem is how to verify the entangled states that are generated from new genuinely multipartite entangled states in the present model by using Bell-type inequalities.

Theorem 3 provides a useful method for verifying new genuinely multipartite entanglement in the present local model by using witness operators \cite{HHH,Chru}. Similar to the biseparable model \cite{Sy}, all the states with the decompositions in Eq.(\ref{eqn1}) consist of a convex set. This feature implies another application of witness operator \cite{HHH,Chru}. In applications, the state tomography will be performed to find proper entangled pure state for constructing witness operator from Theorem 3. Another method is to find an entangled mixed state which closes to the verified state. Generally, we may define a witness operator for verifying $\rho$ as
\begin{eqnarray}
{\cal W}_{\rho}=\zeta\mathbbm{1}-\rho
\label{eqna2}
\end{eqnarray}
where $\zeta=\sup_{\rho_{net}}({\rm tr}[\sqrt{\rho_{net}}\sqrt{\rho}\sqrt{\rho_{net}}])^2$ \cite{Jozsa} and $\rho_{net}$ is any network state defined in Eq.(\ref{eqn1}). Unfortunately, it is difficult to evaluate $\zeta$ for general states. Hence, an interesting problem is how to compute $\zeta$ for special states.

The present Theorems 1 and 3 are useful for featuring permutationally symmetric entangled states. Actually, there is a standard Bell method to verify some of these entangled states. Especially, we show that all the biseparable states $\rho$ given in Eq.(\ref{eq4}) on Hilbert space $\otimes_{i=1}^n\mathbb{H}_{A_i}$ satisfy the following inequality (Supplementary E)
\begin{eqnarray}
&&\sum_{1\leq i\not=j\leq n}\langle{}M_i\otimes{}M_j\rangle_{\rho}-
\frac{n-1}{n}\sum_{i=1}^n(f_1(\rho,M_i)
\nonumber
\\
&&+\sum_{j\not=i}^nf_2(\rho,M_i,M_j)) \leq  n-1
\label{a1}
\end{eqnarray}
where $\langle{}M_i\otimes{}M_j\rangle_{\rho}={\rm tr}[(M_i\otimes{}M_j)\rho]$, $f_1$ and $f_2$ are nonlinear operators defined by $f_1(\rho,M_i)={\rm tr}[\rho^p(M_i\otimes \mathbbm{1})\rho^{1-p}(M_i\otimes \mathbbm{1})]$ and $f_2(\rho,M_i,M_j)={\rm tr}[\rho^{p}(M_i\otimes{}\mathbbm{1})\rho^{1-p} (\mathbbm{1}\otimes{}M_j)]$ with $p\in (0,1)$, and $M_i$s are dichotomic quantum observable. The optimal bound is $n^2-n$ for general states. The interesting feature is that the inequality (\ref{a1}) only requires two-body correlations \cite{TAS}. This inequality is useful for verifying GHZ state \cite{GHZ}, partial W state \cite{Dicke1}, Dicke states \cite{Toth}, and generalized entangled 3-qubit states \cite{AAC} in the biseparable model \cite{Sy} (Supplementary E). Note that from Theorem 2, all the multipartite entangled qubit states in the biseparable model \cite{Sy} are also new genuinely multipartite entangled in the present model given in Eq.(\ref{eqn1}). This means that the inequality (\ref{a1}) provides the first Bell inequality for verifying new genuinely multipartite entangled in the present model given in Eq.(\ref{eqn1}). Unfortunately, it is inefficient for the genuinely multipartite entangled states derived from networks, such as its shown in Fig.\ref{fig-1}(b) and (c). This may be valuable for further investigations.

\section{Conclusion}

In conclusion, we propose a new local model to verify new genuinely multipartite entanglement going beyond those generated by multisource quantum networks with shared randomness and local operations without classical communication. The new genuinely multipartite entanglement is stronger than its being verified in the biseparable model or other models with multiple independent sources. The first result implies a new generic feature of permutationally symmetric entangled pure states in the present local model. Similar result holds for other multipartite entangled pure states in the biseparable model on Hilbert space with local dimensions no larger than three. These results show new generic multipartite nonlocality. Interestingly, the present new genuinely multipartite entanglement is consistent with its verified in the biseparable model for specific systems such as noisy GHZ states and noisy qubit states. The present results are interesting in Bell theory, quantum information processing and measurement-based quantum computation.

After finishing the present manuscript \cite{Luo20}, we became aware of two independent works in ref.\cite{Nav} and ref.\cite{Kraft}. In ref.\cite{Nav}, authors have defined a similar local model under the assumptions of linear operations for local parties. They then prove that a generalized GHZ state and W state are genuinely multipartite entanglement using the inflation technique. Another result is obtained for tripartite GHZ state in ref.\cite{Kraft} by using local unitary operations. Our results have four improvements compared with these proved in refs.\cite{Nav,Kraft}. One is from Theorem 1 which provides a generic result for all the permutationally symmetric entangled pure states. The second is from Theorem 2 which shows a generic feature of all multipartite entangled pure states with small local dimensions. Moreover, it shows that the present model and the biseparable model can be used to verify the same kind of multipartite entangled pure states. The third is from Theorem 3 for witnessing general noisy states. This shows the robustness of new genuinely multipartite entanglement. The last is the inequality (\ref{a1}) for verifying new multipartite nonlocality using only two-body correlations.

\section*{Acknowledgements}

We thank Ronald de Wolf, Carlos Palazuelos, Luming Duan, Yaoyun Shi, Ying-Chang Liang, Donglin Deng, Xiubo Chen, and Yuan Su. This work was supported by the National Natural Science Foundation of China (Nos.61772437,61702427), Sichuan Youth Science and Technique Foundation (No.2017JQ0048), and Fundamental Research Funds for the Central Universities (No.2682014CX095).

\section*{Conflict of Interest}

The authors declare no conflict of interest.

\section*{Keywords}

Quantum entanglement, genuinely multipartite nonlocality, quantum network, permutationally symmetric states, entangled sources

\section*{Supplementary A: Proof of Theorem 1}

In this section, we prove Theorem 1, i.e., any $n$-partite ($n\geq 3$) permutationally symmetric entangled pure state is new genuinely $n$-partite entanglement in the local model given in Eq.(1) in the main text. Note that classical communication is not allowed for our model. From Definition 1 in the main text, it is sufficient to consider the local unitary operations for all parties because a pure state can only be generated by using pure state under local unitary operations. To show the main idea, we firstly prove the following lemmas.

\textbf{Lemma 1}. Consider an $n$-partite permutationally symmetric pure state $|\Phi\rangle_{A_1\cdots{}A_n}$ on Hilbert space $\mathbb{H}_{A_1}\otimes \cdots\otimes\mathbb{H}_{A_n}$ shared by $n$ parties $\textsf{A}_1, \cdots, \textsf{A}_n$. Assume that $|\Phi\rangle$ can be decomposed into two states under the local unitary operations $W_1\otimes \cdots{}\otimes{}W_n$, i.e.,
\begin{align*}
(W_1\otimes \cdots{}\otimes{}W_n)|\Phi\rangle=|\Phi_1\rangle|\Phi_2\rangle
\tag{A1}
\end{align*}
where $|\Phi_1\rangle$ and $|\Phi_2\rangle$ are states on some subspaces of $\mathbb{H}_{A_1}\otimes \cdots\otimes\mathbb{H}_{A_n}$. Then, there are local operation $W$ and two pure states $|\Psi_1\rangle$ and $|\Psi_2\rangle$ on some subspaces of $\mathbb{H}_{A_1}\otimes \cdots\otimes\mathbb{H}_{A_n}$ such that
\begin{align*}
(W\otimes \cdots \otimes W )|\Phi\rangle=|\Psi_1\rangle|\Psi_2\rangle
\tag{A2}
\end{align*}

{\bf Proof of Lemma 1}. The pr oof is from the symmetry of $|\Phi\rangle_{A_1\cdots{}A_n}$. From the assumption in Eq.(A1), there is a local system, $A_1$ for example, which can be decomposed into two subsystems $A_{11}$ and $A_{12}$ under the local operation $W_1$ with $\mathbb{H}_{A_1}=\mathbb{H}_{A_{11}}\otimes \mathbb{H}_{A_{12}}$, i.e.,
\begin{align*}
(W_1\otimes \mathbbm{1}_{A_2\cdots{}A_n})|\Phi\rangle&=(\mathbbm{1}_{A_1}\otimes W_2^{-1}\otimes \cdots \otimes W_n^{-1})|\Phi_1\rangle|\Phi_2\rangle
\\
:&=|\hat{\Phi}_1\rangle_{A_{11}B_1}|\hat{\Phi}_2\rangle_{A_{12}B_2}
\tag{A3}
\end{align*}
where $\mathbbm{1}$ denotes the identity operator on $\mathbb{H}_{A_2}\otimes \cdots \otimes \mathbb{H}_{A_n}$, and $W_i^{-1}$ denotes the inverse matrix of $W_i$, $A_{11}$ may be entangled with one system $B_1$, $A_{12}$ may be entangled with one system $B_2$, two systems $B_1$ and $B_2$ satisfy $\mathbb{H}_{B_1}\otimes \mathbb{H}_{B_2}=\mathbb{H}_{A_2}\otimes \cdots\otimes\mathbb{H}_{A_n}$.

From the symmetry of $|\Phi\rangle_{A_1\cdots{}A_n}$, by using local operation $W_1$ the local system $A_i$ will also be decomposed into two subsystems $A_{i1}$ and $A_{i2}$ with $\mathbb{H}_{A_i}=\mathbb{H}_{A_{i1}}\otimes \mathbb{H}_{A_{i2}}$, $i=2, \cdots, n$. From Eq.(A3), it means that
\begin{align*}
(W_1\otimes \cdots{}\otimes{}W_1)|\Phi\rangle=|\Psi_1\rangle_{A_{11}\cdots A_{n1}}|\Psi_2\rangle_{A_{12}\cdots A_{n2}}
\tag{A4}
\end{align*}
This completes the proof. $\Box$

From Lemma 1 and Eq.(1) in the main text, it is sufficient to prove permutationally symmetric entangled pure state under the same local operation for all parties. The following lemma is used to classify all the possible decompositions of a permutationally symmetric entangled pure state. Here, we take tripartite entangled state as an example.

\textbf{Lemma 2}. Consider a tripartite permutationally symmetric entangled pure state $|\Phi\rangle_{A_1A_2,B_1B_2,C_1C_2}$ in the biseparable model \cite{Sy} on Hilbert space $\mathbb{H}_{A_1}\otimes \mathbb{H}_{A_2}\otimes\mathbb{H}_{B_1}\otimes\mathbb{H}_{B_2}\otimes \mathbb{H}_{C_1}\otimes \mathbb{H}_{C_2}$, where the party $\textsf{A}_1$ has qubits $A_1$ and $A_2$, $\textsf{A}_2$ has qubits $B_1$ and $B_2$, and $\textsf{A}_3$ has qubits $C_1$ and $C_2$. Assume that $|\Phi\rangle_{A_1A_2,B_1B_2,C_1C_2}$ can be decomposed into two states under the local operation $W$, i.e.,
\begin{align*}
(W\otimes W\otimes W)|\Phi\rangle=|\Phi_1\rangle|\Phi_2\rangle
\tag{A5}
\end{align*}
for two pure states $|\Phi_1\rangle$ and $|\Phi_2\rangle$. Then the following results hold:
\begin{itemize}
\item{}\textbf{Decomposition}-We have (under permutations of $A_1$ and $A_2$, $B_1$ and $B_2$, or $C_1$ and $C_2$)
\begin{align*}
(W\otimes W\otimes W)|\Phi\rangle_{A_1A_2,B_1B_2,C_1C_2}&=|\Psi_1\rangle_{A_1B_1C_1}|\Psi_2\rangle_{A_2B_2C_2}
\tag{A6}
\\
(W\otimes W\otimes W)|\Phi\rangle_{A_1A_2,B_1B_2,C_1C_2}&=|\psi\rangle_{A_1}|\psi\rangle_{B_1}|\psi\rangle_{C_1}
|\Psi\rangle_{A_2B_2C_2}
\tag{A7}
\end{align*}
\item{}\textbf{Symmetry}- All the states of $|\Psi\rangle_{A_2B_2C_2}$, $|\Psi_1\rangle_{A_1B_1C_1}$ and $|\Psi_2\rangle_{A_2B_2C_2}$ are permutationally symmetric.
\item{}\textbf{Entanglement}- $|\Psi\rangle_{A_2B_2C_2}$ is a tripartite entanglement in the biseparable model \cite{Sy}, and one of $|\Psi_1\rangle_{A_1B_1C_1}$ and $|\Psi_2\rangle_{A_2B_2C_2}$ is a tripartite entanglement in the biseparable model \cite{Sy}.
\end{itemize}

\textbf{Proof of Lemma 2}. Note that $A_i, B_i, C_i$ are qubits which cannot be further decomposed into the tensor of Hilbert spaces with at least two dimensions.  From Eq.(A5), there are two cases: one is the local systems owned by the parties $\textsf{A}_i$ and $\textsf{A}_{i+1}$ in $(W\otimes W\otimes W)|\Phi\rangle$ are separable for some $i\leq 2$. The other is that $A_1$ and $A_2$ (or $B_1$ and $B_2$, or $C_1$ and $C_2$) in $(W\otimes W\otimes W)|\Phi\rangle$ are separable. For the first case, by using the symmetry of $(W\otimes W\otimes W)|\Phi\rangle$, it follows that all the local systems owned by $\textsf{A}_i$ are separable, $i=1, 2$. Hence, $(W\otimes W\otimes W)|\Phi\rangle$ is a product state, which contradicts to the assumption that $|\Phi\rangle$ is an entanglement. Hence, the only possible decomposition is the second case, i.e., decomposing $A_1$ and $A_2$, or $B_1$ and $B_2$, or $C_1$ and $C_2$.

In what follows, the proof is completed by three steps.

\textbf{Step 1. Proof of decomposition}

Assume that $|\Phi\rangle_{A_1A_2,B_1B_2,C_1C_2}$ has the following decomposition
\begin{align*}
(W\otimes W\otimes W)|\Phi\rangle_{A_1A_2,B_1B_2,C_1C_2}=|\Phi_1\rangle_{A_1B_1}|\Phi_2\rangle_{A_2B_2C_1C_2}
\tag{A8}
\end{align*}
where $|\Phi_i\rangle$s are pure states. Since $|\Phi\rangle_{A_1A_2,B_1B_2,C_1C_2}$ is permutationally symmetric, by swapping the joint systems $A_1A_2$ and $B_1B_2$, we get
\begin{align*}
(W\otimes W\otimes W)|\Phi\rangle_{A_1A_2,B_1B_2,C_1C_2}=|\Phi_1\rangle_{B_1A_1}|\Phi_2\rangle_{B_2A_2C_1C_2}
\tag{A9}
\end{align*}
Similarly, by swapping the joint systems $A_1A_2$ and $C_1C_2$, we get
\begin{align*}
(W\otimes W\otimes W)|\Phi\rangle_{A_1A_2,B_1B_2,C_1C_2}=|\Phi_1\rangle_{C_1B_1}|\Phi_2\rangle_{C_2B_2A_1A_2}
\tag{A10}
\end{align*}
From Eq.(A9), $B_1$ and $B_2$ in $(W\otimes W\otimes W)|\Phi\rangle_{A_1A_2,B_1B_2,C_1C_2}$ are separable. From Eq.(A11), $C_1$ and $C_2$ in $(W\otimes W\otimes W)|\Phi\rangle_{A_1A_2,B_1B_2,C_1C_2}$ are separable. From Eq.(A8), we have one of the following decompositions
\begin{align*}
(W\otimes W\otimes W)|\Phi\rangle_{A_1A_2,B_1B_2,C_1C_2}=|\Phi_1\rangle_{A_1B_1}
|\Phi_{21}\rangle_{A_2C_1}|\Phi_{22}\rangle_{B_2C_2}
\tag{A11}
\end{align*}
and
\begin{align*}
(W\otimes W\otimes W)|\Phi\rangle_{A_1A_2,B_1B_2,C_1C_2}=|\Phi_1\rangle_{A_1B_1}
|\Phi_{21}\rangle_{A_2C_2}|\Phi_{22}\rangle_{B_2C_1}
\tag{A12}
\end{align*}
for some states $|\Phi_{21}\rangle$ and $|\Phi_{22}\rangle$.

For the decompositions in Eqs.(A11) and (A12),  $C_1$ and $B_1$ are separable for $|\Phi_1\rangle_{C_1B_1}$ in Eq.(A11), i.e., we have
\begin{align*}
|\Phi_1\rangle_{C_1B_1}=|\phi_{11}\rangle_{C_1}|\phi_{12}\rangle_{B_1}
\tag{A13}
\end{align*}
for some states $|\phi_{11}\rangle$ and $|\phi_{12}\rangle$. Moreover, from Eq.(A11) and (A13), $C_1$ and $A_2$ are separable for $|\Phi_{21}\rangle_{A_2C_1}$ in Eq.(A11), i.e., we have
\begin{align*}
|\Phi_{21}\rangle_{A_2C_1}=|\phi_{211}\rangle_{A_2}|\phi_{212}\rangle_{C_1}
\tag{A14}
\end{align*}
for some states $|\phi_{211}\rangle$ and $|\phi_{212}\rangle$. From Eq.(A8) and (A13), $A_1$ and $B_1$ are separable for $|\Phi_{1}\rangle_{A_1B_1}$ in Eq.(A8), i.e., we have
\begin{align*}
|\Phi_{1}\rangle_{A_2C_1}=|\phi_{11}\rangle_{A_1}|\phi_{12}\rangle_{B_1}
\tag{A15}
\end{align*}
From Eqs.(A11), and (A13)-(A15), $A_1,B_1, C_1, A_2, B_2$ and $C_2$ are separable in $(W\otimes W\otimes W)|\Phi\rangle_{A_1A_2,B_1B_2,C_1C_2}$. It means that $(W\otimes W\otimes W)|\Phi\rangle_{A_1A_2,B_1B_2,C_1C_2}$ is a product state which contradicts to the assumption that $(W\otimes W\otimes W)|\Phi\rangle_{A_1A_2,B_1B_2,C_1C_2}$ is a tripartite entanglement in the biseparable model \cite{Sy}.

Similar results hold for the decomposition in Eq.(A12). This implies that the decomposition in Eq.(A8) is impossible under local operations.

Now, assume that one can decompose one qubit from a permutationally symmetry pure state using local operation $W$, i.e.,
\begin{align*}
(W\otimes W\otimes W)|\Phi\rangle_{A_1A_2,B_1B_2,C_1C_2}=|\psi\rangle_{A_1}|\Phi'\rangle_{A_2B_1B_2C_1C_2}
\tag{A16}
\end{align*}
for some states $|\psi\rangle_{A_1}$ and $|\Phi'\rangle_{A_2B_1B_2C_1C_2}$. Since $(W\otimes W\otimes W)|\Phi\rangle_{A_1A_2,B_1B_2,C_1C_2}$ is permutationally symmetric,  $A_{1}$ should be permutationally symmetric with some qubits $B_i$ and $C_j$. Hence, from the symmetry of $(W\otimes W\otimes W)|\Phi\rangle_{A_1A_2,B_1B_2,C_1C_2}$, we have
\begin{align*}
(W\otimes W\otimes W)|\Phi\rangle_{A_1A_2,B_1B_2,C_1C_2}=|\psi\rangle_{A_1}|\psi\rangle_{B_i}|\psi\rangle_{C_j}
|\Phi''\rangle_{A_2B_{i'}C_{j'}}
\tag{A17}
\end{align*}
for some state $|\Phi''\rangle_{A_2B_{i'}C_{j'}}$, where $i'$ $(i'\in \{0,1\})$ is different from $i$, and $j' (j'\in \{0,1\})$ is different from $j$. Hence, one can decompose three qubits under local operations, i.e., the decomposition in Eq.(A7).

For other case, one can decompose three qubits by using local operations as the decomposition in Eq.(A6).

\textbf{Step 2. Proof of symmetry}

Since $|\Phi\rangle_{A_1A_2,B_1B_2,C_1C_2}$ is a tripartite permutationally symmetric state, $(W\otimes W\otimes W)|\Phi\rangle_{A_1A_2,B_1B_2,C_1C_2}$ is a tripartite permutationally symmetric state. From Eq.(A7), it follows that $|\Psi\rangle_{A_2B_2C_2}$ is permutationally symmetric.

Now, we will prove that $|\Psi_1\rangle_{A_1B_1C_1}$ and $|\Psi_2\rangle_{A_2B_2C_2}$ are permutationally symmetric. Note that $A_i$, $B_j$ and $C_k$ are all qubit systems. It means that $A_1$ and $B_i,C_j$ should be symmetric in the state $|\Psi_1\rangle_{A_1B_1C_1}|\Psi_2\rangle_{A_2B_2C_2}$ for some $i,j$. There are four subcases as follows.
\begin{itemize}
\item[(1)] If $i=j=1$, it follows that $|\Psi_1\rangle_{A_1B_1C_1}$ is permutationally symmetric. Moreover, $|\Psi_2\rangle_{A_2B_2C_2}$ is also permutationally symmetric because $|\Psi_1\rangle_{A_1B_1C_1}|\Psi_2\rangle_{A_2B_2C_2}$ is tripartite permutationally symmetric.
\item[(2)] If $i=j=2$, the proof is similar to the case (1).

\item[(3)] If $i=1$ and $j=2$, it follows from Eq.(A6) that $B_1$ and $C_2$ are separable. Hence, $A_1$ and $B_1$ should also separable in Eq.(A6). We get
\begin{align*}
|\Psi_1\rangle_{A_1B_1C_1}|\Psi_2\rangle_{A_2B_2C_2}
=|\psi\rangle_{A_1}|\psi\rangle_{B_1}|\psi\rangle_{C_2}|\phi\rangle_{C_1}|\Psi_3\rangle_{A_2B_2}
\tag{A18}
\end{align*}
for some qubit states $|\psi\rangle$ and $|\phi\rangle$,  and two-qubit state $|\Psi_3\rangle_{A_2B_2}$. However, the right side of Eq.(A18) is at most bipartite entanglement which contradicts to the assumption that $|\Phi\rangle_{A_1A_2,B_1B_2,C_1C_2}$ is a tripartite entanglement in the biseparable model \cite{Sy}. Hence, this case is impossible.
\item[(4)] If $i=2$ and $j=1$, it is also impossible for Eq.(A6). The proof is similar to the case (3).
\end{itemize}
To sum up, we get both $|\Psi_1\rangle_{A_1B_1C_1}$ and $|\Psi_2\rangle_{A_2B_2C_2}$ are tripartite permutationally symmetric.

\textbf{Step 3. Proof of entanglement}

Since $|\Phi\rangle_{A_1A_2,B_1B_2,C_1C_2}$ is a tripartite entanglement in the biseparable model \cite{Sy}, from the decomposition in Eq.(A7) $|\Psi\rangle_{A_2B_2C_2}$ is a tripartite entanglement in the biseparable model \cite{Sy}. Moreover, $|\Psi_1\rangle_{A_1B_1C_1}|\Psi_2\rangle_{A_2B_2C_2}$ in Eq.(A6) is a tripartite entanglement in the biseparable model \cite{Sy}, it follows that one of $|\Psi_1\rangle_{A_1B_1C_1}$ and $|\Psi_2\rangle_{A_2B_2C_2}$ is a tripartite entanglement in the biseparable model \cite{Sy}, where the joint systems $A_1A_2$, $B_1B_2$ and $C_1C_2$ are owned by different parties. The proof is completed by contradiction. In fact, assume that both $|\Psi_1\rangle_{A_1B_1C_1}$ and $|\Psi_2\rangle_{A_2B_2C_2}$ are bipartite entangled states. It means that $|\Psi_1\rangle_{A_1B_1C_1}$ and $|\Psi_2\rangle_{A_2B_2C_2}$ are separable states. Since $|\Psi_1\rangle_{A_1B_1C_1}$ and $|\Psi_2\rangle_{A_2B_2C_2}$ are permutationally symmetric, we get that $|\Psi_1\rangle_{A_1B_1C_1}$ and $|\Psi_2\rangle_{A_2B_2C_2}$ are product states. This contradicts to the assumption that $|\Phi\rangle_{A_1A_2,B_1B_2,C_1C_2}$ is a tripartite entanglement in the biseparable model \cite{Sy}. This completes the proof. $\Box$

Now, similar to Lemma 2, we can get all the possible decompositions of $n$-partite permutationally symmetric pure state.

\textbf{Lemma 3}. Consider an $n$-partite permutationally symmetric entangled pure state $|\Phi\rangle_{A_1B_1,\cdots, A_nB_n}$ in the biseparable model \cite{Sy} shared by $n$ parties $\textsf{A}_1,\cdots, \textsf{A}_n$, where the party $\textsf{A}_i$ has qubits $A_i$ and $B_i$, $i=1, \cdots, n$. If $|\Phi\rangle_{A_1B_1,\cdots,A_nB_n}$ can be decomposed into two states, i.e.,
\begin{align*}
(W\otimes \cdots\otimes W)|\Phi\rangle_{A_1B_1,\cdots,A_nB_n}=|\Phi_1\rangle|\Phi_2\rangle
\tag{A19}
\end{align*}
the following results hold
\begin{itemize}
\item{}\textbf{Decomposition}-We have (under permutations of $A_i$ and $B_i$, $i=1, \cdots, n$)
\begin{align*}
(W\otimes \cdots \otimes W)|\Phi\rangle_{A_1B_1,\cdots,A_nB_n}&=|\Psi_1\rangle_{A_1\cdots A_n}|\Psi_2\rangle_{B_1\cdots B_n}
\tag{A20}
\\
(W\otimes \cdots \otimes W)|\Phi\rangle_{A_1B_1,\cdots,A_nB_n}&=|\psi\rangle_{A_1}\otimes \cdots\otimes|\psi\rangle_{A_n}
|\Psi\rangle_{B_1\cdots{}B_n}
\tag{A21}
\end{align*}
\item{}\textbf{Symmetry}-All the states $|\Psi\rangle_{B_1\cdots{}B_n}$, $|\Psi_1\rangle_{A_1\cdots{}A_n}$ and $|\Psi_2\rangle_{B_1\cdots{}B_n}$ are permutationally symmetric.
\item{}\textbf{Entanglement}-$|\Psi\rangle_{B_1\cdots{}B_n}$ is an $n$-partite entanglement in the biseparable model \cite{Sy}, and one of $|\Psi_1\rangle_{A_1\cdots{}A_n}$ and $|\Psi_2\rangle_{B_1\cdots{}B_n}$ is an $n$-partite entanglement in the biseparable model \cite{Sy}.
\end{itemize}

\textbf{Proof of Lemma 3}. The proof is similar to its for Lemma 2 with three steps as follows.

\textbf{Step 1. Proof of Decompositions}

From Eq.(A21), there are two cases: one is the local systems owned by $\textsf{A}_i$ and $\textsf{A}_{i+1}$ in $(W\otimes \cdots\otimes W)|\Phi\rangle$ are separable for some $i\leq n-1$. The other is that two qubits of $A_i$ and $B_i$ in the state $(W\otimes \cdots\otimes W)|\Phi\rangle$ are separable. For the first case, by using the symmetry of $(W\otimes \cdots\otimes W)|\Phi\rangle$, it follows that all the joint systems $A_iB_i$ and $A_{j}B_{j}$ are separable. It follows that $(W\otimes \cdots\otimes W)|\Phi\rangle$ is a product state, which contradicts to the assumption that $(W\otimes \cdots\otimes W)|\Phi\rangle$ is an $n$-partite entanglement in the biseparable model \cite{Sy}. Hence, the only possible case in Eq.(A19) is to decompose the local qubits of some party. In what follows, take $A_1$ and $B_1$ as an example, i.e., $A_1$ and $B_1$ in $(W\otimes \cdots\otimes W)|\Phi\rangle_{A_1B_1,\cdots,A_nB_n}$ are separable. From the symmetry of $(W\otimes \cdots\otimes W)|\Phi\rangle_{A_1B_1,\cdots,A_nB_n}$, by swapping the joint systems $A_1B_1$ and $A_iB_i$, we obtain from Eq.(A19) that $A_i$ and $B_i$ in $(W\otimes \cdots\otimes W)|\Phi\rangle_{A_1B_1,\cdots,A_nB_n}$  are separable, where $(W\otimes \cdots\otimes W)|\Phi\rangle_{A_1B_1,\cdots,A_nB_n}$ is invariant under swapping the joint systems  $A_1B_1$ and $A_iB_i$, $i=1, \cdots, n$. It means that $A_i$ and $B_i$ are separable in $(W\otimes \cdots\otimes W)|\Phi\rangle_{A_1B_1,\cdots,A_nB_n}$ for all $i=2, \cdots, n$. Hence, from Eq.(A19), all the possible decompositions (under permutations of $A_i$ and $B_i$, $i=1, \cdots, n$) are shown in Eq.(A20) or Eq.(A21).

\textbf{Step 2. Proof of Symmetry}

Since $|\Phi\rangle_{A_1B_1,\cdots,A_nB_n}$ is an $n$-partite permutationally symmetric state, it follows that $(W\otimes \cdots\otimes W)|\Phi\rangle_{A_1B_1,\cdots,A_nB_n}$ is also an $n$-partite permutationally symmetric state. Hence, from the decomposition in Eq.(A20), $|\Psi\rangle_{B_1\cdots{}B_n}$ is an $n$-partite permutationally symmetric state.

Since $|\Phi\rangle_{A_1B_1,\cdots,A_nB_n}$ is an $n$-partite permutationally symmetric entanglement in the biseparable model \cite{Sy}, we have from Eq.(A20) that $|\Psi_1\rangle_{A_1\cdots{}A_n}|\Psi_2\rangle_{B_1\cdots{}B_n}$ is an $n$-partite permutationally symmetric entanglement in the biseparable model \cite{Sy}. Note that $A_i$ and $B_j$ are qubit systems. It means that $A_1$ and $A_{i}$ (or $A_1$ and $B_i$) should be permutationally symmetric in $|\Psi_1\rangle_{A_1\cdots{}A_n}|\Psi_2\rangle_{B_1\cdots{}B_n}$ for $i=2, \cdots, n$. There are two subcases as follows.
\begin{itemize}
\item[(1)] If all $A_{i}$s are permutationally symmetric, it follows that both $|\Psi_1\rangle_{A_1\cdots{}A_n}$  and $|\Psi_2\rangle_{B_1\cdots{}B_n}$ are permutationally symmetric.
\item[(2)] If $A_{i}$ and $B_{j}$ are permutationally symmetric for some $i\not=j$,  from the decomposition in Eq.(A20), it follows that $A_{i}$ and $B_{j}$ are separable in the right side of Eq.(A20). Take $i=n-1$ and $j=n$ as an example, i.e., $A_1, \cdots, A_{n-1}, B_n$ are permutationally symmetric. We get from Eq.(A20) that
\begin{align*}
|\Psi_1\rangle_{A_1\cdots A_n}|\Psi_2\rangle_{B_1\cdots B_n}
=|\Psi_3\rangle_{A_1\cdots A_{n-1}}|\psi_1\rangle_{A_n}|\psi_2\rangle_{B_n}|\Psi_4\rangle_{B_1\cdots B_{n-1}}
\tag{A22}
\end{align*}
for two qubit states $|\psi_1\rangle_{A_n}$ and $|\psi_2\rangle_{B_n}$, and two $n-1$-qubit states $|\Psi_3\rangle_{A_1\cdots A_{n-1}}$ and $|\Psi_4\rangle_{B_1\cdots B_{n-1}}$. However, the right side of Eq.(A22) is at most $n-1$-partite entanglement in the biseparable model \cite{Sy}, which contradicts to the assumption that $|\Phi\rangle_{A_1B_1,\cdots,A_nB_n}$ is an $n$-partite entanglement in the biseparable model \cite{Sy}. Hence, this case is impossible. Similar proof holds for other cases of $i$ and $j$.
\end{itemize}

To sum up, we have proved that both $|\Psi_1\rangle_{A_1\cdots{}A_n}$ and $|\Psi_2\rangle_{B_1\cdots{}B_n}$ are permutationally symmetric.

\textbf{Step 3. Proof of Entanglement}

Since $|\Phi\rangle_{A_1B_1,\cdots,A_nB_n}$ is an $n$-partite entanglement in the biseparable model \cite{Sy}, we get that $|\Psi\rangle_{B_1\cdots{}B_n}$ is an $n$-partite entanglement in the biseparable model \cite{Sy} from Eq.(A21), and one of $|\Psi_1\rangle_{A_1\cdots{}A_n}$ and $|\Psi_2\rangle_{B_1\cdots{}B_n}$ is an $n$-partite entanglement in the biseparable model \cite{Sy}.

Note that in Eq.(A20), $|\Psi_1\rangle_{A_1\cdots{}A_n}|\Psi_2\rangle_{B_1\cdots{}B_n}$ is an $n$-partite entanglement. It follows that one of $|\Psi_1\rangle_{A_1\cdots{}A_n}$ and $|\Psi_2\rangle_{B_1\cdots{}B_n}$ is an $n$-partite entanglement in the biseparable model \cite{Sy}. The proof is completed by contradiction. In fact, assume that both $|\Psi_1\rangle_{A_1\cdots{}A_n}$ and $|\Psi_2\rangle_{B_1\cdots B_n}$ are $n-1$-partite entangled states in the biseparable model \cite{Sy}. It means that $|\Psi_1\rangle_{A_1\cdots{}A_n}$ and $|\Psi_2\rangle_{B_1\cdots B_n}$ are separable states. Since $|\Psi_1\rangle_{A_1\cdots{}A_n}$ and $|\Psi_2\rangle_{B_1\cdots B_n}$ are permutationally symmetric, we get that $|\Psi_1\rangle_{A_1\cdots{}A_n}$ and $|\Psi_2\rangle_{B_1\cdots B_n}$ are product states, which contradicts to the assumption that $|\Phi\rangle_{A_1B_1,\cdots,A_nB_n}$ is an $n$-partite entanglement in the biseparable model \cite{Sy}. This completes the proof. $\Box$

Note that the decomposition in Eq.(A21) is special case of Eq.(A20). Hence, in what follows, we only need to consider the decomposition in Eq.(A20).

{\bf Proof of Theorem 1}. Consider an $n$-partite permutationally symmetric entangled pure state $|\Phi\rangle_{A_1\cdots{} A_n}$ in the biseparable model \cite{Sy} on Hilbert space $\otimes_{i=1}^n\mathbb{H}_{A_i}$, where $d={\rm dim}(\mathbb{H}_{A_i})$ satisfying $d\geq 2$, $i=1, \cdots, n$. Note that $|\Phi\rangle$ cannot be decomposed into a linear superposition of two different mixed states $\rho_1$ and $\rho_2$, i.e., $|\Phi\rangle\langle \Phi|\not=p\rho_1+(1-p)\rho_2$ for any probability distribution $\{p, 1-p\}$ with $p\not=0,1$. From Definition 1 in the main text, it is sufficient to prove that $|\Phi\rangle$ cannot be generated from any pure state under local unitary operations.

The proof is completed by contradiction. Assume that there is an $n$-partite permutationally symmetric entangled state $|\Phi\rangle_{A_{1},\cdots, A_n}$ in the biseparable model \cite{Sy} which can be decomposed into two entangled states under the local unitary operations $W_i$s, i.e.,
\begin{align*}
(W_1\otimes \cdots \otimes W_n)|\Phi\rangle_{A_{1},\cdots, A_n}=|\Phi_1\rangle|\Phi_2\rangle
\tag{A23}
\end{align*}
where both $|\Phi_1\rangle$ and $|\Phi_2\rangle$ are at most $n-1$-partite entangled states in the biseparable model \cite{Sy}. From Lemma 1, it follows that
\begin{align*}
(W\otimes \cdots \otimes W)|\Phi\rangle_{A_{1},\cdots, A_n}=|\hat{\Phi}_1\rangle|\hat{\Phi}_2\rangle
\tag{A24}
\end{align*}
for some local operation $W$ and two states $|\hat{\Phi}_1\rangle$ and $|\hat{\Phi}_2\rangle$, where both $|\hat{\Phi}_1\rangle$ and $|\hat{\Phi}_2\rangle$ are at most $n-1$-partite entangled states in the biseparable model \cite{Sy}.

In what follows, we construct Algorithm 1 to prove Theorem 1. The main idea is to iteratively decompose $|\Phi\rangle_{A_{1},\cdots, A_n}$ into its defined in Eq.(A20). Assume that $\mathbb{H}_{A_i}$ is $d$-dimensional space. By embedding $d$-dimensional space $\mathbb{H}_{d}$ into $2^k$-dimensional space $\mathbb{H}_{2^k}$ with $d\leq 2^k$, i.e., $|i\rangle_{A_i}\mapsto|i_1\cdots i_d\rangle_{A_{i1}\cdots{}A_{id}}$ with binary representation $i_1\cdots i_d$ of $i$, $i=1, \cdots, d$, it is sufficient to consider all the qubit systems for each party. Here, assume that $|\Phi\rangle$ is written into $|\Phi\rangle_{A_{11}\cdots{}A_{1k},\cdots, A_{n1}\cdots{}A_{nk}}$ on Hilbert space $\mathbb{H}_{A_{11}}\otimes\cdots\otimes \mathbb{H}_{A_{nk}}$, where the party $\textsf{A}_i$ has $k$ qubits $A_{i1}, \cdots, A_{ik}$, $i=1, \cdots, n$.

\begin{center}
\begin{algorithm}
{\bf Algorithm 1}
\begin{itemize}
\item[Input] An $n$-partite permutationally symmetric entangled state $|\Phi\rangle_{A_{11}\cdots{}A_{1k},\cdots, A_{n1}\cdots{}A_{nk}}$ in the biseparable model \cite{Sy}, where $A_{ij}$s are qubit systems. Define $|\Phi^{(0)}\rangle_{A^{(0)}_1,\cdots, A^{(0)}_n}:=|\Phi\rangle_{A_{11}\cdots{}A_{1k},\cdots, A_{n1}\cdots{}A_{nk}}$, where $\mathbb{H}_{A^{(0)}_i}=\otimes_{j=1}^k\mathbb{H}_{A_{ij}}$.

\item[\textsf{For}] $s=1$ to $s=k$

\item[] \% Applying Lemma 3.
    \begin{itemize}
 \item[\textsf{if}] the decomposition in Eq.(A20) is possible for $|\Phi^{(s-1)}\rangle_{A^{(s-1)}_1,\cdots, A^{(s-1)}_n}$

 \item[]From Eq.(A20), there is a new $n$-partite permutationally symmetric entangled state $|\Phi^{(s)}\rangle_{A^{(s)}_1,\cdots, A^{(s)}_n}$ in the biseparable model \cite{Sy} on Hilbert space $\mathbb{H}_{A^{(s)}_1}\otimes\cdots\otimes \mathbb{H}_{A^{(s)}_n}$, where
    \begin{align*}
    \mathbb{H}_{A^{(s)}_i}\subset \mathbb{H}_{A^{(s-1)}_i}, i=1, \cdots, n
    \tag{A25}
    \end{align*}
    i.e., for each party the state space of $|\Phi^{(s)}\rangle_{A^{(s)}_1,\cdots, A^{(s)}_n}$ is smaller than its of $|\Phi^{(s-1)}\rangle_{A^{(s-1)}_1,\cdots, A^{(s-1)}_n}$.
\item[\textsf{else}]
\item[] $|\Phi^{(s-1)}\rangle_{A^{(s-1)}_1,\cdots, A^{(s-1)}_n}$ cannot be decomposed into two pure states under any local unitary operations.
\item[\textsf{return}] An $n$-partite $|\Phi^{(s-1)}\rangle_{A^{(s-1)}_1,\cdots, A^{(s-1)}_n}$
\item[\textsf{endif}]
\end{itemize}
\item[\textsf{Output}] An $n$-partite $|\Phi^{(t)}\rangle_{A^{(t)}_1,\cdots, A^{(t)}_n}$
\end{itemize}
\end{algorithm}
\end{center}

There are two facts in Algorithm 1. One is that if the decomposition in Eq.(A20) is applicable for input state $|\Phi^{(s)}\rangle_{A^{(s)}_1,\cdots, A^{(s)}_n}$, from Lemma 3, we get a new $n$-partite permutationally symmetric entangled state $|\Phi^{(s)}\rangle_{A^{(s+1)}_1,\cdots, A^{(s+1)}_n}$ on Hilbert space $\otimes_{j=1}^n\mathbb{H}_{A^{(s+1)}_j}$ which is smaller than the space $\otimes_{j=1}^n\mathbb{H}_{A^{(s)}_j}$ from Eq.(A25). This implies that the decomposition in Algorithm 1 will end in finite iterations. The other is that the output is an $n$-partite permutationally symmetric entangled state $|\Phi^{(s-1)}\rangle_{A^{(s-1)}_1,\cdots, A^{(s-1)}_n}$ or $|\Phi^{(t)}\rangle_{A^{(t)}_1,\cdots, A^{(t)}_n}$ in the biseparable model \cite{Sy} from Lemma 3. It means that after all the possible decompositions, there is at least one $n$-partite permutationally symmetric entanglement in the biseparable model \cite{Sy}. This contradicts to the decomposition in Eq.(A24), where all the decomposed states are at most $n-1$-partite entanglement in the biseparable model \cite{Sy}. Hence, any $n$-partite permutationally symmetric entangled state in the biseparable model \cite{Sy} cannot be decomposed into the states in Eq.(1) in the main text. This completes the proof.

\section*{Supplementary B: Proof of Theorem 2}

In the following proof, any isomorphism mapping (including the embedding mapping which maps one Hilbert space into a subspace of larger Hilbert space, see example in the proof of Theorem 1) is allowed for Hilbert space $\mathbb{H}_{A_i}$. One important fact is that these isomorphism mappings do not change the number of parties in any entangled system, even if isomorphism mappings may change the number of particles in some entangled system. This allows that all isomorphism mappings can be performed before all local unitary operations (or encoded into local unitary operations assisted by auxiliary systems).

{\bf Lemma 4}. Consider two multipartite entangled pure states $|\Psi_1\rangle_{A_1\cdots{}A_k}$ and $|\Psi_2\rangle_{A_k'A_{k+1}\cdots A_n}$ in the biseparable model \cite{Sy}. The particles of $A_k$ and $A'_{k}$ are entangled after a unitary operation $U_k$ being performed if $U_k\not=U_{A_k}\otimes W_{A_k'}$ for any unitary operations $U_{A_k}$ and $W_{A_k'}$, where $U_{A_k}$ and $W_{A_k'}$ are performed on the respective system $A_k$ and $A_k'$.

{\bf Proof of Lemma 4}. The Schmidt decomposition of $|\Psi_1\rangle$ and $|\Psi_2\rangle$ are given by
\begin{align*}
&|\Psi_1\rangle=\sum_{i=0}^{d-1}\alpha_i
|\psi_i\rangle_{A_1\cdots{}A_{k-1}}
|\tau_i\rangle_{A_k}
\tag{B1}
\\
&|\Psi_2\rangle=\sum_{i=0}^{d-1}\beta_i
|\tau_i\rangle_{A_k'}
|\phi_i\rangle_{A_{k+1}\cdots{}A_{n}}
\tag{B2}
\end{align*}
where $\{|\tau_i\rangle\}$ are orthogonal states of $A_k'$, $\{|\psi_i\rangle_{A_1\cdots{}A_{k-1}}\}$ are $k-1$-partite orthogonal states of $A_1, \cdots{}, A_{k-1}$, and $\{|\phi_i\rangle_{A_{k+1}\cdots{}A_{n}}\}$ are $n-k$-partite orthogonal states of $A_{k+1}, \cdots{}, A_{n}$. Since $|\Psi_i\rangle$s are entangled states in the biseparable model \cite{Sy}, there are at least two nonzero Schmidt coefficients in Eqs.(B1) and (B2). Here, we choose $\{|\tau_i\rangle\}$ as the basis of $A_k$ or $A_k'$. Otherwise, proper local unitary operation can change different bases into the same basis states. Consider a unitary operation $U_k$ satisfying $U_k\not=U_{A_k}\otimes W_{A_k'}$ for any unitary operations $U_{A_k}$ and $W_{A_k'}$. Note that
\begin{align*}
|\Omega\rangle:=&(\mathbbm{1}_{r}\otimes U_k)|\Psi_1\rangle|\Psi_2\rangle
\\
=&
\sum_{i,j}\alpha_i\beta_j|\varsigma_{ij}\rangle
|\phi_i\rangle|\psi_j\rangle
\tag{B3}
\end{align*}
where $\mathbbm{1}_{r}$ denotes the identity operator on all the systems $A_i$s except for the systems $A_k$ and $A_k'$, and  $\{|\varsigma_{ij}\rangle, |\varsigma_{ij}\rangle=U_k|ij\rangle\}$ are orthogonal states for all $i,j$.

Assume that the systems $A_k$ and $A_k'$ of $|\Omega\rangle$ defined in Eq.(B3) is not entangled in the biseparable model \cite{Sy}. The bipartition $\{A_1, \cdots, A_{k}\}$ and $\{A_{k+1}, \cdots, A_{n}\}$ (or $\{A_1, \cdots, A_{k-1}\}$ and $\{A_k', A_{k+1}, \cdots, A_{n}\}$) is also not entangled for $|\Omega\rangle$. It follows that the bipartition $\{A_1, \cdots, A_{k}\}$ and $\{A_{k}', A_{k+1}, \cdots, A_{n}\}$ is also not entangled for $|\Omega\rangle$ in Eq.(B3). It means that $|\Omega\rangle$ is a product state of two subsystems $\{A_1, \cdots, A_{k}\}$ and $\{A_{k}', A_{k+1}, \cdots, A_{n}\}$ in the biseparable model \cite{Sy}, i.e.,
\begin{align*}
|\Omega\rangle&=|\Psi'_1\rangle_{A_1 \cdots A_{k}}
|\Psi'_2\rangle_{A_{k}' A_{k+1} \cdots A_{n}}
\\
&=\sum_{i,j}\alpha_i\beta_j(\mathbbm{1}_r\otimes{}U_{A_k}\otimes{}W_{A_{k}'})
|\tau_{i}\rangle|\tau_j\rangle
|\phi_i\rangle|\psi_j\rangle
\tag{B4}
\end{align*}
where $|\Psi'_1\rangle_{A_1 \cdots A_{k}}=\sum_{i}\alpha_i(\mathbbm{1}_{A_1\cdots{}A_{k-1}}\otimes{}U_{A_k})|\tau_i\rangle
|\phi_i\rangle$ and $|\Psi'_2\rangle_{A_1 \cdots A_{k}}=\sum_{i}\gamma_i(\mathbbm{1}_{A_{k+1}\cdots{}A_{n}}\otimes{}W_{A_k'})|\tau_i\rangle|\psi_i\rangle$
for some single-particle unitary operations $U_{A_k}$ and $W_{A_k'}$ since the subsystem of $A_1, \cdots, A_{k-1}$ of $|\Psi'_i\rangle_{A_1\cdots A_{k}}$ and $|\Psi'_i\rangle_{A_1\cdots A_{k}}$ are the same to each other, $\mathbbm{1}_{A_1\cdots{}A_{k-1}}$ denotes the identity operator on all the systems $A_1,\cdots{}, A_{k-1}$ and $\mathbbm{1}_{A_{k+1}\cdots{}A_{n}}$ denotes the identity operator on all the systems $A_{k+1},\cdots{}, A_{n}$.

It follows from Eqs.(B3) and (B4) that
\begin{align*}
|\varsigma_{ij}\rangle=(U_{A_k}\otimes{}W_{A_{k}'})
|\tau_{i}\rangle_{A_k}|\tau_j\rangle_{A_{k}'}
\tag{B5}
\end{align*}
for all $i, j$. Note that $U_k|\tau_{i}\rangle_{A_k}|\tau_j\rangle_{A_{k}'}=|\varsigma_{ij}\rangle$ for any $i,j$. It follows that $U_k=U_{A_k}\otimes{}W_{A_{k}'}$ since $\{|\tau_{i}\rangle|\tau_j\rangle\}$ are the orthogonal states. This contradicts to the assumption that  $U_k$ is not the tensor of two local unitary operations. This completes the proof. $\square$

Similarly, we can prove the following lemma.

\textbf{Lemma 5}. Consider a $k$-partite entangled pure state $|\Omega\rangle$ on Hilbert space $\otimes_{i=1}^n\mathbb{H}_{A_i}$. $|\Omega\rangle$ is also $k$-partite entangled after any local unitary operations.

\textbf{Lemma 6}. Each Hilbert space $\mathbb{H}$ with dimension $d\leq 3$ cannot be decomposed into the tensor of two Hilbert spaces $\mathbb{H}_1$ and $\mathbb{H}_2$ with at least two dimensions.

\textbf{Proof of Lemma 6}. Assume that $\mathbb{H}$ can be decomposed into the tensor of two Hilbert spaces $\mathbb{H}_1, \mathbb{H}_2$, i.e., $\mathbb{H}= \mathbb{H}_1\otimes \mathbb{H}_2$, where $\mathbb{H}_1$ and $\mathbb{H}_2$ have at least two dimensions. In this case, there are at least four orthogonal basis states in $\mathbb{H}_1\otimes \mathbb{H}_2$. The number of basis states dose not decrease under any local unitary operations because Hilbert space $\mathbb{H}_1\otimes \mathbb{H}_2$ is not isomorphic to Hilbert space $\mathbb{H}$ with dimension smaller than $4$. This contradicts to the assumption that $\mathbb{H}$ has local dimension $d\leq 3$. $\Box$.

\textbf{Proof of Theorem 2}. Consider an entangled pure state $|\Phi\rangle$ on Hilbert space $\otimes_{i=1}^n\mathbb{H}_{A_i}$ with local dimensions $d_1, \cdots, d_n\leq 3$. Note that any pure state cannot be decomposed into a superposition of two mixed states, where the classical communication is not allowed for all parties. It is sufficient to consider pure states in Definition 1 in the main text.

The proof is completed by induction of the number $\ell$ of decomposed states in Eq.(1) in the main text.

{\bf Case one}. $\ell=2$, i.e., $|\Phi\rangle$ cannot be generated by two entangled states $|\Psi_1\rangle$ and $|\Psi_2\rangle$ under local unitary operations, where $|\Psi_1\rangle$ and $|\Psi_2\rangle$ are at most $n-1$-partite entangled states.

 Consider any two multipartite entangled states $|\Psi_1\rangle$ and $|\Psi_2\rangle$ in the biseparable model \cite{Sy}. For simplicity, assume that $|\Psi_1\rangle$ is on Hilbert space $\otimes_{i=1}^k\mathbb{H}_{A_i}$ and $|\Psi_2\rangle$ is on Hilbert space $\otimes_{j=k}^n\mathbb{H}_{A_j}$. Here, we assume that the $k$-th party is the middle party who shares two entangled states $|\Psi_1\rangle$ and $|\Psi_2\rangle$.  Otherwise, $|\Psi_1\rangle|\Psi_2\rangle$ is a separable state which cannot be locally transformed into $|\Phi\rangle$ on Hilbert space $\otimes_{i=1}^n\mathbb{H}_{A_i}$ with local unitary operations from Lemma 5.

In what follows, we prove that all the parties cannot get $|\Phi\rangle$ from $|\Psi_1\rangle|\Psi_2\rangle$ deterministically using local unitary operations.

Assume that there exist local operations $U_i$s satisfying that
\begin{align*}
&(\otimes_{i=1}^nU_i)|\Psi_1\rangle_{A_1\cdots{}A_k}
|\Psi_2\rangle_{A'_kA_{k+1}\cdots{}A_n}
=|\Phi\rangle_{A_1A_2\cdots{}A_n}|\phi\rangle_{A'_k}
\tag{B6}
\end{align*}
where $|\phi\rangle_{A'_k}$ is an auxiliary state of $A_k$. Otherwise, $A_k$ and $A'_k$ are entangled after local operations $\otimes_{i=1}^nU_i$ on $|\Psi_1\rangle_{A_1\cdots{}A_k}
|\Psi_2\rangle_{A'_kA_{k+1}\cdots{}A_n}$. It means that the joint system of $A_k$ and $A'_k$ in $|\Psi_1\rangle_{A_1\cdots{}A_k}
|\Psi_2\rangle_{A'_kA_{k+1}\cdots{}A_n}$ have at least four orthogonal basis states because both $A_k$ in $|\Psi_1\rangle_{A_1\cdots{}A_k}$ and $A_k'$ in $|\Psi_2\rangle_{A'_kA_{k+1}\cdots{}A_n}$ have at least two orthogonal basis states. It means that for the $i$-th party the joint systems $A_k$ and $A'_k$ in $(\otimes_{i=1}^nU_i)|\Psi_1\rangle_{A_1\cdots{}A_k}
|\Psi_2\rangle_{A'_kA_{k+1}\cdots{}A_n}$ have at least four orthogonal basis states. This implies that the state space of the $i$-l party is at least four dimensions, which contradicts to the assumption of $d_j\leq 3$ for $i=1, \cdots, n$.

Note that the local operations of $A_i$s except for $A_k$ can be performed any time because there is no classical communication. Hence, it is reasonable to assume that the local operations of all parties except for the $k$-th party are included in the states $|\Psi_1\rangle$ and $|\Psi_2\rangle$. In this case, Eq.(B6) is rewritten into
\begin{align*}
(\mathbbm{1}_r\otimes U_k)|\Psi_1\rangle_{A_1\cdots{}A_k}
|\Psi_2\rangle_{A_k'A'_{k+1}\cdots{}A_n}
=|\Phi\rangle_{A_1A_2\cdots{}A_n}|\phi\rangle_{A'_k}
\tag{B7}
\end{align*}
where $U_{k}$ is performed on the joint system of $A_k$ and $A'_{k}$.

If $U_k\not=U_{A_k}\otimes W_{A_k'}$ for any unitary operations $U_{A_k}$ and $W_{A_k'}$, from Lemma 4 the subsystems of $A_k$ and $A'_{k}$ are entangled. Note that local unitary operations do not change the entanglement of other systems. It means that $A_1, \cdots, A_{k-1}$ are entangled, and $A_{k+1}, \cdots, A_{n}$ are entangled after $U_{A_k}$ being performed. From Lemma 5, it follows that $A_1, \cdots, A_{k}$ are entangled, and $A_{k'}, \cdots, A_{n}$ are entangled in the biseparable model \cite{Sy}. Hence, $A_1, \cdots, A_k, A_{k'}, \cdots, A_{n}$ are entangled in the biseparable model \cite{Sy}. It means that $(\mathbbm{1}_r\otimes U_k)|\Psi_1\rangle_{A_1\cdots{}A_k}$ is an $n+1$-partite entanglement in the biseparable model \cite{Sy}. This contradicts to the product state of $|\Phi\rangle_{A_1A_2\cdots{}A_n}|\phi\rangle_{A'_k}$ in the right side of Eq.(B7).

Similar results hold for the case that there are $s$ parties who share two entangled states $|\Psi_1\rangle$ and $|\Psi_2\rangle$ simultaneously.

{\bf Case two}. Assume that $|\Phi\rangle$ cannot be generated by no more than $\ell=m-1$ entangled states $|\Psi_1\rangle, \cdots, |\Psi_{\ell}\rangle$  under local unitary operations, where $|\Psi_1\rangle, \cdots, |\Psi_{m-1}\rangle$ are at most $n-1$-partite entangled states in the biseparable model \cite{Sy}. In what follows, we prove the result for $\ell=m$.

Note that an $n$-partite entangled pure state $|\Phi\rangle$ in the biseparable model \cite{Sy} cannot be generated by an $n-1$-partite entangled pure state $|\Psi\rangle$ and any separable states by local unitary operations from Lemma 5. Assume that there are $m$ states $|\Psi_1\rangle, \cdots, |\Psi_m\rangle$ for generating $|\Phi\rangle$, i.e.,
\begin{align*}
(\otimes_{i=1}^nU_i)
(\otimes_{j=1}^m|\Psi_j\rangle)
=|\Phi\rangle|\Phi_0\rangle
\tag{B8}
\end{align*}
where $|\Psi_i\rangle$s are at most $n-1$-partite entangled states in the biseparable model \cite{Sy}, $U_i$s are unitary operations on local system of $i$-th party, $i=1, \cdots, n$, and $|\Phi_0\rangle$ is an auxiliary state. $\mathbb{H}_{A_i}$ cannot be decomposed into the tensor of two Hilbert spaces, it is sufficient to assume that the dimensions of all the particles involved in $|\Psi_j\rangle$s are no larger than three. Otherwise, we can decompose a large Hilbert space into the tensor of small Hilbert spaces.

Now, define $|\Gamma\rangle_{B_1\cdots{}B_s}=\otimes_{j=1}^{m-1}|\Psi_j\rangle$. From the assumption, $|\Gamma\rangle$ cannot be used to generate $|\Phi\rangle$ by using local operations. Here, we cannot apply Case one above because $|\Gamma\rangle$ may be an $n$-partite entanglement. Assume that $|\Gamma\rangle$ is a $t$-partite entanglement in the biseparable model \cite{Sy}. If $t\leq n-1$, from Case one, $|\Gamma\rangle$ and $|\Psi_m\rangle$ cannot be used to generate $|\Phi\rangle$ by using local operations. This contradicts to Eq.(B8). It means that the result holds for $\ell=m$. It completes the proof.

If $t=n$, we have $d_i\geq 2$, where $d_i$ denotes the dimension of $\mathbb{H}_{B_i}$, $i=1, \cdots, t$. From Lemma 5, $|\Phi_0\rangle$ should be generated by using local unitary operations on $\otimes_{j=1}^m|\Psi_j\rangle$. Consider the joint system of
\begin{align*}
|\Omega_{ij}\rangle=|\Phi_i\rangle|\Phi_j\rangle
\tag{B9}
\end{align*}
for any $i\not=j$.
\begin{itemize}
\item{}If $|\Omega_{ij}\rangle$ is at most $n-1$-partite entanglement for some $i\not=j$, $\otimes_{j=1}^m|\Psi_j\rangle$ can be regarded as a tensor of $m-1$ states $|\Psi_k\rangle$ with $k\not=i,j$, and $|\Omega_{ij}\rangle$. Eq.(B8) contradicts to the assumption of $\ell=m-1$, i.e., any $m-1$ states $|\Psi_1\rangle, \cdots, |\Psi_{m-1}\rangle$ cannot be used to generate $|\Phi\rangle$ under local operations. It means that the result holds for $\ell=m$. This completes the proof.
\item{}All the states of $|\Omega_{ij}\rangle$s are $n$-partite entangled states for $i\not=j$ in the biseparable model \cite{Sy}. In this case, from Lemma 5, any local operations cannot be used to create new entanglement from separable states. Moreover, from Eq.(B8), the local state spaces of $|\Psi_j\rangle$s cannot be decomposed into the tensor of two Hilbert spaces with at least two dimensions. Hence, the entanglement $|\Phi_0\rangle$ in Eq.(B8) should be generated by $|\Psi_i\rangle$s, i.e.,
\begin{align*}
|\Phi_0\rangle=(W_{1}\otimes \cdots \otimes{}W_t)(\otimes_{i\in I}|\Psi_i\rangle)
\tag{B10}
\end{align*}
for some local operations $W_i$, where $I\subset\{1, \cdots, n\}$. Combined Eq.(B8) and (B10), we have
\begin{align*}
|\Phi\rangle=(\otimes_{i=1}^nU_i)(W^{-1}_{1}\otimes \cdots W^{-1}_t)(\otimes_{j\in \overline{I}}|\Psi_j\rangle
\tag{B11}
\end{align*}
where $\overline{I}$ denotes the complement set of $I$ in the set $\{1, \cdots, n\}$. From Eq.(B11), $|\Phi\rangle$ can be generated by $|\overline{I}|$ (with $|\overline{I}|\leq n-1$) entangled states which are at most $n-1$-partite entangled states in the biseparable model \cite{Sy}. This contradicts to the assumption of $\ell=m-1$.
\end{itemize}
It means that the result holds for $\ell=m$. This completes the proof.

\section*{Supplementary C: Proof of Theorem 3}

Before we prove Theorem 3, we firstly prove the following lemma.

{\bf Lemma 7}. Consider Hilbert space $\otimes_{i=1}^n\mathbb{H}_{A_i}$, where $\mathbb{H}_{A_i}$ has dimension $d_i$ satisfying $d_i\geq 2$, $i=1, \cdots, n$. The following inequalities hold
\begin{align*}
&D(|GHZ\rangle)=\max\{a^2_0,\cdots, a^2_{d-1}\}
\tag{C1}
\\
&D(|D_{k,n}\rangle)=D(|D_{nd-n-1-k,n}\rangle)=\frac{n-1}{n+k-1}, k=1, \cdots, \lfloor\frac{nd-n-1}{2}\rfloor
\tag{C2}
\\
&D(|\Phi_{ps}\rangle)\leq \alpha_0^2\beta^2+
\sum_{i=1}^{\lfloor\frac{nd-n-1}{2}\rfloor}\frac{n-1}{n+i-1}(\alpha_i^2+\alpha_{nd-n-1-i}^2),
\mbox{ for odd } nd-n-1
\tag{C3}
\\
&D(|\Phi_{ps}\rangle)\leq \alpha_0^2\beta^2+
\sum_{i=1}^{\lfloor\frac{nd-n-1}{2}\rfloor}\frac{n-1}{n+i-1}(\alpha_i^2+\alpha_{nd-n-1-i}^2)
-\frac{2n-2}{nd+n-3}\alpha_{(nd-n-1)/2}^2
\mbox{ for even } nd-n-1
\tag{C4}
\\
&D(|\Phi\rangle)=\max_{{\cal A}\subseteq\{A_1,\cdots, A_n\}}\max\{\sigma(\rho_{\cal A})\}, \mbox{ for } d_1= \cdots=d_n=2
\tag{C5}
\end{align*}
where $|GHZ\rangle$ is generalized GHZ state \cite{GHZ} defined in Eq.(3) in the main text, $|D_{k,n}\rangle$ are Dicke states \cite{Dicke1} defined in Eq.(4) in the main text, $|\Phi_{ps}\rangle$ is a generalized permutationally symmetric entangled state defined in Eq.(12) in the main text, $\beta$ in Eq.(C3) or (C4) is given by $\beta=\max\{\beta_0,\beta_1\}$, $\sigma(\rho_{\cal A})$ denotes all the eigenvalues of $\rho_{\cal A}$, and $\rho_{\cal A}$ denotes the reduced density matrix $\rho_{\cal A}$ of the subsystems in ${\cal A}$ with ${\cal A}\subset\{A_1, \cdots, A_n\}$.

The proof of Theorem 3 is easily followed from Lemma 7 and the inequality (8) in the main text.

{\bf Proof of Lemma 7}. The proof is completed by several steps as follows.

We firstly show that the definition of $D(|\Phi\rangle)$ in the main text is reasonable even if we do not evaluate $D(|\Phi\rangle|\Phi_0\rangle)$ with an axillary state $|\Phi_0\rangle$.

Consider an $n$-partite entangled pure state $|\Phi\rangle$ on Hilbert space $\otimes_{i=1}^n\mathbb{H}_{A_i}$. Here, we consider that $|\Phi\rangle$ is permutationally symmetric,  or qubit state with $d_i=2$ for $i=1, \cdots, n$. From the definition of $D(|\Phi\rangle)$ in the main text, for an auxiliary state $|\Phi_0\rangle$ on Hilbert space $\otimes_{j=1}^m\mathbb{H}_{B_j}$, consider any pure state $|\Omega\rangle$ on Hilbert space $(\otimes_{i=1}^n\mathbb{H}_{A_i})\otimes (\otimes_{j=1}^m\mathbb{H}_{B_j})$, which is generated by the states $|\Psi_1\rangle, \cdots, |\Psi_t\rangle$, where all $|\Psi_i\rangle$s are at most $n-1$-partite entangled in the biseparable model \cite{Sy}. From Definition 1, $|\Omega\rangle$ is given by $|\Omega\rangle:=\otimes_{j=1}^nU_j
(\otimes_{i=1}^t|\Psi_i\rangle)$ for some local unitary operations $U_1, \cdots, U_n$. Now, define the Schmidt decomposition of $|\Omega\rangle_{A_1\cdots{}A_nB_1\cdots{}B_m}$ in terms of the bipartition $\{A_1, \cdots, A_n\}$ and $\{B_1, \cdots, B_m\}$ as
\begin{align*}
|\Omega\rangle_{A_1\cdots{}A_nB_1\cdots{}B_m}=\sum_{i=0}^s\lambda_i
|\Omega_i\rangle_{A_1\cdots{}A_n}
|\Phi_i\rangle_{B_1\cdots{} B_m}
\tag{C6}
\end{align*}
where $\{|\Omega_0\rangle, \cdots, |\Omega_s\rangle\}$ are orthogonal states on Hilbert space $\otimes_{i=1}^n\mathbb{H}_{A_i}$, $\{|\Phi_0\rangle, \cdots, |\Phi_s\rangle\}$ are orthogonal states on Hilbert space $\otimes_{j=1}^n\mathbb{H}_{B_j}$, and $\lambda_i$s are Schmidt coefficients satisfying $0\leq \lambda_i\leq 1$ and $\sum_{i=0}^s\lambda_i^2=1$. Here, we assume the first basis state of $\otimes_{j=1}^n\mathbb{H}_{B_j}$ is $|\Phi_0\rangle$. Otherwise, we can change it into
$|\Phi_0\rangle$ by using a local operation on $\otimes_{j=1}^n\mathbb{H}_{B_j}$. From Eq.(6) in the main text, it follows that
\begin{align*}
D(|\Phi\rangle|\Phi_0\rangle)&=\sup_{|\Omega\rangle,|\Phi_0\rangle}
|\langle \Omega |\Phi\rangle|\Phi_0\rangle|^2
\\
&= \sup_{|\Omega_0\rangle,\lambda_0}
\lambda_0^2|\langle \Omega_0 |\Phi\rangle|^2
\\
&\leq \sup_{|\Omega_0\rangle}|\langle \Omega_0 |\Phi\rangle|^2
\\
&=D(|\Phi\rangle)
\tag{C7}
\end{align*}
It means that the maximal distance of $D(|\Phi\rangle)$ is achieved by exploring the maximal distance between the network state defined in Eq.(1) in the main text and the new genuinely multipartite entanglement $|\Phi\rangle$ in the present model. It does not need to consider a generalized distance $D(|\Phi\rangle|\Phi_0\rangle)$ with any axillary state $|\Phi_0\rangle$. Hence, the definition of $D(|\Phi\rangle)$ in the main text is reasonable.

In what follows, we prove the inequalities (C1)-(C5).

{\bf Case 1. Proof of Inequality (C1)}

To show the main idea, we firstly prove the result for a special $n$-partite GHZ state as
\begin{align*}
|\Phi\rangle_{A_1\cdots{}A_n}=a_0|0\rangle^{\otimes n}+a_{d-1}|d-1\rangle^{\otimes n}
\tag{C8}
\end{align*}
on Hilbert space $\otimes_{i=1}^n\mathbb{H}_{A_i}$, where all the spaces  $\mathbb{H}_{A_i}$ have the same dimension $d$ with $d\geq 2$, $|x\rangle^{\otimes n}$ denotes the tensor of $n$ number of $|x\rangle$, and $a_0^2+a_{d-1}^2=1$. Note that $|\Phi\rangle$ can be regarded as a projection on the subspace spanned by $\{|0\rangle^{\otimes n}, |d-1\rangle^{\otimes n}\}$.

For two decomposed states $|\Psi_0\rangle$ and $|\Psi_1\rangle$ of $|\Omega_0\rangle$ in Eq.(C7), i.e., $|\Omega_0\rangle=|\Psi_0\rangle|\Psi_1\rangle$, we assume that
\begin{align*}
|\Psi_0\rangle&=\sum_{i=0}^{d-1}
x_{i}|i\rangle_{A_k'}|\Psi'_{i}\rangle_{A_1\cdots{}A_{k-1}}
\tag{C9}
\\
|\Psi_1\rangle&=\sum_{j=0}^{d-1}y_j|j\rangle_{A_k''}
|\Psi''_j\rangle_{A_{k+1}\cdots{}A_n}
\tag{C10}
\end{align*}
In Eq.(C9), $|\Psi_i'\rangle$s are orthogonal states on Hilbert space $\mathbb{H}_{A_1}\otimes \cdots{}\otimes\mathbb{H}_{A_{k-1}}$ satisfying $x_i\geq 0$ and  $\sum_{i=0}^{d-1}x_i^2=1$. In Eq.(C10), $|\Psi_j''\rangle$ are orthogonal states on Hilbert space $\mathbb{H}_{A_{k+1}}\otimes \cdots{}\otimes\mathbb{H}_{A_{n}}$ satisfying $y_i\geq0$, $\sum_{i=0}^{d-1}y_i^2=1$. Here, the $k$-th party shares two states $|\Psi_0\rangle$ and $|\Psi_1\rangle$.

Note that $\{x_iy_j\}$ are Schmidt coefficients of $|\Psi_0\rangle|\Psi_1\rangle$ with bipartition of $\{A_k'A_k''\}$ and $\{A_1$, $\cdots$, $A_{k-1},A_{k+1}$, $\cdots, A_n\}$. These Schmidt coefficients are not changed under local unitary operations of all parties, where one can re-change the orthogonal basis of the joint system $A_k'A_k''$, or equivalently the orthogonal basis of $A_k$ in $|\Phi\rangle$. From Eq.(C8), there are only two Schmidt coefficients $x_i^2y_j^2$ and $x_l^2y_s^2$ which are useful to construct $|\Phi\rangle$. In this case, we assume that $|\Psi_i\rangle$s have the following decompositions as
\begin{align*}
|\Psi_0\rangle&=x_0|0\rangle^{\otimes m_1}+x_1|d-1\rangle^{\otimes m_1}+x_2|\Psi'_0\rangle
\tag{C11}
\\
|\Psi_1\rangle&=y_0|0\rangle^{\otimes m_2}+y_1|d-1\rangle^{\otimes m_2}+y_2|\Psi'_1\rangle
\tag{C12}
\end{align*}
with $m_i\leq n-1$, $i=1, 2$. In Eq.(C11), $|\Psi_0\rangle$ is on Hilbert space $\otimes_{i=1}^{m_1}\mathbb{H}_{A_i}$, $x_i\geq0$ and $\sum_{i}x_i^2=1$. $|\Psi'_0\rangle$ is a state which is orthogonal to the subspace spanned by $\{|0\rangle^{\otimes m_1}, |d-1\rangle^{\otimes m_1}\}$. In Eq.(C12), $|\Psi_1\rangle$ is on Hilbert space $\otimes_{i=1}^{m_1}\mathbb{ H}_{B_i}$, $y_i\geq 0$ and $\sum_{i}y_i^2=1$. $|\Psi'_1\rangle$ is a state which is orthogonal to the subspace spanned by $\{|0\rangle^{\otimes m_2}, |d-1\rangle^{\otimes m_2}\}$. Hilbert space $\otimes_{i=1}^n\mathbb{ H}_{A_i}$ is isomorphic to $(\otimes_{i=1}^{m_1}\mathbb{H}_{A_i})\otimes(\otimes_{j=1}^{m_1}\mathbb{ H}_{B_j})$.

From Eqs.(C7), (C11) and (C12) we get that
\begin{align*}
|\langle \Phi|\Psi_0\rangle|\Psi_1\rangle|^2
=& (a_0x_0y_0+x_1y_1a_{d-1})^2
\\
\leq &\max_{\sum_{i=0}^{d-1}x_i^2\leq 1\atop{\sum_{j=0}^{d-1}y_j^2\leq 1}}(a_0x_0y_0+x_1y_1a_{d-1})^2
\\
\leq & a_0^2-a_0^2(x_0^2+y_0^2)+(a_0^2+a_{d-1}^2)x_0^2y_0^2
+2a_0a_{d-1}x_0y_0\sqrt{(1-x_0^2)(1-y_0^2)}
\tag{C13}
\\
\leq & a_0^2-2a_0^2x_0y_0+x_0^2y_0^2
+2a_0a_{d-1}x_0y_0(1-x_0y_0)
\tag{C14}
\\
=&(1-2a_0a_{d-1})x_0^2y_0^2
+(2a_0a_{d-1}-2a_0^2)x_0y_0+a_0^2
\\
\leq &\max\{a_0^2,a_{d-1}^2\}
\tag{C15}
\end{align*}
Inequality (C13) follows from the inequalities $x_0^2+x_1^2\leq 1$ and $y_0^2+y_1^2\leq 1$. Inequality (C14) follows from the equality: $a_0^2+a_{d-1}^2=1$ and inequality: $x_0^2+y_0^2\geq 2x_0y_0$. Inequality (C15) follows from the fact that $f(x_0y_0)=(a_0^2-2a_0a_{d-1})x_0^2y_0^2+(2a_0a_{d-1}
-2a_0^2)x_0y_0+a_0^2$
is a quadratic function of $x_0y_0$ with $1-2a_0a_{d-1}\geq 0$ because $2a_0a_{d-1}\leq a_0^2+a_{d-1}^2=1$. Thus the maximum is achievable at $x_0y_0=0$ or $x_0y_0=1$.

Consider a new decomposition of $|\Omega_0\rangle$ in Eq.(C7) as
\begin{align*}
|\Omega_0\rangle=\otimes_{i=0}^k|\Psi_i\rangle
\tag{C16}
\end{align*}
where $|\Psi_i\rangle=x_{0,i}|0\rangle^{\otimes m_i}+x_{1,i}|d-1\rangle^{\otimes m_i}+x_{2,i}|\Psi'_{0,i}\rangle$, $i=1, \cdots, k$. It follows that
\begin{align*}
|\langle \Phi|\otimes_{i=0}^k|\Psi_i\rangle|^2&= (a_0\prod_{j=0}^kx_{0,j}+a_{d-1}\prod_{j=0}^kx_{1,j})^2
\\
&\leq \max_{\sum_{j=0}^{d-1}x_{i,j}^2\leq 1}
(a_0\prod_{j=0}^kx_{0,j}+a_{d-1}\prod_{j=0}^kx_{1,j})^2
\\
&\leq \max_{\sum_{j=0}^{d-1}x_{i,j}^2\leq 1}
(a_0x_{0,0}x_{0,1}+a_{d-1}x_{1,0}x_{1,1})^2
\tag{C17}
\\
&\leq \max\{a_0^2,a_{d-1}^2\}
\tag{C18}
\end{align*}
Inequality (C17) follows from the inequalities $0\leq x_{i,j}\leq 1$ for all $i,j$. Inequality (C18) follows from the inequality (C13). Since Eq.(C18) is achievable, it follows from Eqs.(C6) and (C18) that
\begin{align*}
D(|\Phi\rangle)&=\max\{a_0^2,a_{d-1}^2\}
\tag{C19}
\end{align*}
This completes the proof.

Now, we prove the inequality (C1). Consider a generalized GHZ state as
\begin{align*}
|\Phi\rangle_{A_1\cdots{}A_n}=|GHZ\rangle=\sum_{i=0}^{d-1}a_i|i\rangle^{\otimes n}
\tag{C20}
\end{align*}
$|\Phi\rangle$ can be regarded as a projection onto the subspace spanned by $\{|0\rangle^{\otimes n},\cdots, |d-1\rangle^{\otimes n}\}$. Similar to the discussions from Eq.(C9) to Eq.(C12),  for two decomposed states $|\Psi_0\rangle$ and $|\Psi_1\rangle$ of $|\Omega_0\rangle$ in Eq.(C7), i.e., $|\Psi_0\rangle|\Psi_1\rangle=|\Omega_0\rangle$, it is sufficient to assume that
\begin{align*}
|\Psi_0\rangle&=\sum_{i=0}^{d-1}x_i|i\rangle^{\otimes m_1}+x_*|\Psi'_0\rangle
\tag{C21}
\\
|\Psi_1\rangle&=\sum_{i=0}^{d-1}y_i|i\rangle^{\otimes m_2}+y_*|\Psi'_1\rangle
\tag{C22}
\end{align*}
In Eq.(C21), $|\Psi_0\rangle$ is on Hilbert space $\otimes_{j=1}^{m_1}\mathbb{H}_{A_j}$, $x_i\geq0$ and $x_*^2+\sum_{i}x_i^2=1$. $|\Psi'_0\rangle$ is a state which is orthogonal to the subspace spanned by $\{|0\rangle^{\otimes m_1}, \cdots, |d-1\rangle^{\otimes m_1}\}$. In Eq.(C22), $|\Psi_1\rangle$ is on Hilbert space $\otimes_{j=1}^{m_1}\mathbb{H}_{B_j}$, $y_i\geq0$ and $y_*^2+\sum_{i}y_i^2=1$. $|\Psi'_1\rangle$ is orthogonal to the subspace spanned by $\{|0\rangle^{\otimes m_2}, \cdots, |d-1\rangle^{\otimes m_2}\}$. Hilbert space $\otimes_{i=1}^n\mathbb{H}_{A_i}$ is isomorphic to $(\otimes_{i=1}^{m_1}\mathbb{H}_{A_i})\otimes(\otimes_{j=1}^{m_1}\mathbb{H}_{B_j})$.

From Eqs. (C21) and (C22) we get that
\begin{align*}
|\langle \Phi|\Psi_0\rangle|\Psi_1\rangle|^2 &= (\sum_{i=0}^{d-1}a_ix_iy_i)^2
\\
&\leq \max_{\sum_{i=0}^{d-1}x_i^2\leq 1,\sum_{i=0}^{d-1}y_i^2\leq 1}(\sum_{i=0}^{d-1}a_ix_iy_i)^2
\\
&\leq \max_{\sum_{i=0}^{d-1}x_i^2\leq 1}(\sum_{i=0}^{d-1}a_ix_i^2)^2
\tag{C23}
\\
&\leq \max\{a_0^2, \cdots, a_{d-1}^2\}
\tag{C24}
\end{align*}
Inequality (C23) is obtained by using the following Lagrange method \cite{Ber}:
\begin{align*}
\max_{x_0, \cdots, x_{d-1}\atop{y_0, \cdots, y_{d-1}}}f(x_0, \cdots, x_{d-1}, y_0, \cdots, y_{d-1})
=&(\sum_{i=0}^{d-1}a_ix_iy_i)^2
-\gamma_1(\sum_{i=0}^{d-1}x_i^2-1)
-\gamma_2(\sum_{i=0}^{d-1}y_i^2-1)
\tag{C25}
\\
{\rm s. t.,} &\sum_{i=0}^{d-1}x_i^2-1=0,
\tag{C26}
\\
&\sum_{i=0}^{d-1}y_i^2-1=0,
\tag{C27}
\end{align*}
where $\gamma_1$ and $\gamma_2$ are Lagrange factors. From the equality of $\nabla_{x_i}f=0$ (partial derivative of function $f$ on the variable $x_i$) and Eqs.(C26) and (C27), we get that the maximum of $f(x_0, \cdots, x_{d-1}, y_0, \cdots, y_{d-1})$ achieves when $x_i=y_i$ for $i=0, 1, \cdots, d-1$. Inequality (C24) follows the inequality: $\sum_{i=0}^{d-1}a_ix_i^2\leq \max\{a_0, \cdots, a_{d-1}\}\sum_{i}x_i^2\leq\max\{a_0, \cdots, a_{d-1}\}$.

Moreover, similar to Eqs.(C17) and (C18), the inequality (C24) holds for the decomposition of $\otimes_{j=1}^m|\Psi_j\rangle=|\Omega_0\rangle$ in Eq.(C7). The equalities in Eqs.(C23) and (C24) are achievable. It implies from Eq.(C7) that
\begin{align*}
D(|\Phi\rangle)&=\max\{a_0^2, \cdots, a_{d-1}^2\}
\tag{C28}
\end{align*}

{\bf Case 2. Proof of Inequality (C2)}

Consider a Dicke state as
\begin{align*}
|\Phi\rangle_{A_1\cdots{}A_n}=|D_{k,n}\rangle=
\sum_{x_1+\cdots{}+x_n=k}\frac{1}{\sqrt{N_{k,n}}}
|x_1\cdots{}x_n\rangle
\tag{C29}
\end{align*}
Note that $|D_{k,n}\rangle$ is equivalent to $|D_{nd-n-1-k,n}\rangle$ under the local unitary transformation: $|k\rangle\mapsto |d-k\rangle$, $k=0, \cdots, d-1$, for each party. Hence, we have $D(|D_{k,n}\rangle)=D(|D_{n-k,n}\rangle)$ for $k=0, \cdots, \lfloor\frac{nd-n-1-k}{2}\rfloor$.

In what follows, we need to consider the case of $k\leq \lfloor\frac{nd-n-1-k}{2}\rfloor$.

Consider the Schmidt decomposition of $|\Phi\rangle$ with the bipartition $\{A_1, \cdots, A_s\}$ and $\{A_{s+1}, \cdots, A_n\}$ as
\begin{align*}
|\Phi\rangle=\sum_{i=0}^{\ell}\lambda_i
|\phi_i\rangle|\psi_i\rangle
\tag{C30}
\end{align*}
where $\lambda_i\geq0$ and $\sum_{i}\lambda_i^2=1$,
all the states $|\phi_i\rangle$s are orthogonal states on Hilbert space $\otimes_{i=1}^s\mathbb{H}_{A_i}$, all the states $|\psi_i\rangle$s are orthogonal states on Hilbert space $\otimes_{j=s+1}^n\mathbb{H}_{A_{j}}$. Here, $|\phi_i\rangle$s and $|\psi_i\rangle$s can be chosen as Dicke states. $|\Phi\rangle$ can be regarded as a projection onto the subspace spanned by $\{|\phi_i\rangle|\psi_i\rangle, \forall i\}$. From Theorem 1 in the main text, any $m$-partite Dicke states cannot be generated by using network states (at most $m-1$-partite entangled states) defined in Eq.(1) in the main text. Hence, similar to the discussions from Eq.(C9) to (C12), for the decomposition of $|\Omega_0\rangle=|\Psi_0\rangle|\Psi_1\rangle$ in Eq.(C7), it is sufficient to assume that
\begin{align*}
|\Psi_0\rangle&=\sum_{i=0}^{\ell}x_i|\phi_i\rangle+x_*|\Psi'_0\rangle
\tag{C31}
\\
|\Psi_1\rangle&=\sum_{i=0}^{\ell}y_i|\psi_i\rangle+y_*|\Psi'_1\rangle
\tag{C32}
\end{align*}
In Eq.(C31), $|\Psi_0\rangle$ is on Hilbert space $\otimes_{i=1}^s\mathbb{H}_{A_i}$,  $x_i\geq 0$ and $x_*^2+\sum_{i}x_i^2=1$. $|\Psi'_0\rangle$ is orthogonal to the subspace spanned by $\{|\phi_i\rangle, \forall i\}$. In Eq.(C32), $|\Psi_1\rangle$ is on Hilbert space $\otimes_{j=s+1}^n\mathbb{ H}_{A_{j}}$, $y_i\geq 0$ and $y_*^2+\sum_{i}y_i^2=1$, and $|\Psi'_1\rangle$ is a state which is orthogonal to the subspace spanned by $\{|\psi_i\rangle, \forall i\}$.

From Eqs. (C30)-(C32) we get that
\begin{align*}
|\langle \Phi|\Psi_0\rangle|\Psi_1\rangle|^2 &= (\sum_{i=0}^{\ell}\lambda_ix_iy_i)^2
\\
&\leq \max_{\sum_{i=0}^{\ell}x_i^2\leq 1\atop{\sum_{i=0}^{\ell}y_i^2\leq 1}}(\sum_{i=0}^{\ell}\lambda_ix_iy_i)^2
\\
&\leq \max_{\sum_{i=0}^{\ell}x_i^2\leq 1}(\sum_{i=0}^{\ell}a_ix_i^2)^2
\tag{C33}
\\
&\leq \max\{\lambda_0^2, \cdots, \lambda_{\ell}^2\}
\tag{C34}
\end{align*}
Inequality (C33) follows from the Lagrange method as its stated in Eqs.(C25)-(C27). Inequality (C34) is similar to Eq.(C24).

Since $|\Phi\rangle$ is permutationally symmetric, all the decomposed states $|\phi_i\rangle$ and $|\psi_j\rangle$ are also permutationally symmetric. In this case, we can choose special orthogonal states as
\begin{align*}
&|\phi_i\rangle=|D_{i,s}\rangle,
\\
&|\psi_i\rangle=|D_{k-i,n-s}\rangle,
\tag{C35}
\end{align*}
$i=0, \cdots, k$, where $|D_{l,m}\rangle$ denotes $s$-particle Dicke state with $l$ excitations, i.e, $|D_{l,m}\rangle=\frac{1}{\sqrt{N_{l,m}}}
\sum_{r_1+\cdots+r_m=l}|r_1\cdots{}r_m\rangle$. It means that $|\phi_i\rangle$s and $|\psi_i\rangle$s are also permutationally symmetric states which can be further decomposed with Dicke basis states. From Theorem 1 in the main text, similar to the discussions from Eq.(C9) to (C12), for $m$ decomposed states $|\Psi_1\rangle, \cdots, |\Psi_m\rangle$ of $|\Omega_0\rangle$ in Eq.(C7), it is sufficient to decompose all $|\Psi_i\rangle$s with Dicke basis states. Similar to Case 1, we get that
\begin{align*}
|\langle \Phi|\otimes_{i=1}^m|\Psi_i\rangle|^2
\leq & \max\{\lambda_0^2, \cdots, \lambda_{\ell}^2\}
\tag{C36}
\end{align*}

Note that the maximum in Eq.(C36) is achievable. From the symmetry of $|\Phi\rangle$, it follows from Eqs.(C6) and (C36) that
\begin{align*}
D(|\Phi\rangle)&=\max_{\{A_1, \cdots, A_s\}\cup \{A_{s+1}, \cdots, A_n\}}\max\{\lambda_0^2, \cdots, \lambda_{\ell}^2\}
\tag{C37}
\end{align*}

From Eqs.(C29) and (C35) we get that
\begin{align*}
\lambda_i=\sqrt{\frac{N_{i,s}N_{k-i,n-s}}{N_{k,n}}}, i=0, \cdots, k
\tag{C38}
\end{align*}
From Eqs.(C36) and (C38), it follows that
\begin{align*}
D(|\Phi\rangle)=&\max_{s=1, \cdots, \lfloor{}n/2\rfloor}\max\{
\frac{N_{i,s}N_{k-i,n-s}}{N_{k,n}}, i=0, \cdots, k\}
\\
=&\max_{s=1,  \cdots, \lfloor{}n/2\rfloor}
\max\{
\frac{C(i+s-1,s-1)C(n+k-s-i-1,n-s-1)}{C(n+k-1,n-1)}, i=0, \cdots, k\}
\tag{C39}
\\
=&\max_{s=1, \cdots, \lfloor{}n/2\rfloor}
\max\{
\frac{(i+s-1)!(k+n-i-s-1)!(n-1)!k!}{
(n+k-1)!i!(s-1)!(n-s-1)!(k-i)!}
, i=0, \cdots, k\}
\tag{C40}
\\
=&\max_{s=1, \cdots, \lfloor{}n/2\rfloor}\max\{\frac{(n+k-s-1)!(n-1)!}{
(n+k-1)!(n-s-1)!},
\frac{(k+s-1)!(n-1)!}{
(n+k-1)!(s-1)!}
\}
\tag{C41}
\\
=&\max\{\frac{n-1}{n+k-1}, \frac{(k+\lfloor{}n/2\rfloor-1)!(n-1)!}{
(n+k-1)!(\lfloor{}n/2\rfloor-1)!}\}
\tag{C42}
\\
=&\frac{n-1}{n+k-1}
\tag{C43}
\end{align*}
where $\lfloor{}\frac{n}{2}\rfloor$ denotes the maximal integer no more than $\frac{n}{2}$. In Eq.(C39) we have used the equality: $N_{s,t}=C(s+t-1,t-1)$, which denotes the combination number of choosing $t-1$ ball from a box with $s+t-1$ balls. In Eq.(C30) we have used the equality: $C(\ell_1,\ell_2)=\frac{\ell_1!}{\ell_2!(\ell_1-\ell_2!)}$, $\ell_t!=\prod_{j=1}^{\ell_t}j$. To obtain Eq. (C41) we firstly define $h(i)$ as
\begin{align*}
h(i)=\frac{(i+s-1)!(k+n-i-s-1)!(n-1)!k!}{
(n+k-1)!i!(s-1)!(n-s-1)!(k-i)!}
\tag{C44}
\end{align*}
Note that
\begin{align*}
\frac{h(i+1)}{h(i)}=\frac{(i+s)(k-i)}{(i+1)(k+n-i-s-1)}
\tag{C45}
\end{align*}
which implies that $h(i)$ is a convex function of $i$ with $i\leq k$. Hence, $\max_ih(i)=\max\{h(0),h(k)\}$. Moreover, it is easy to prove that $\varphi(s)=\frac{(n+k-s-1)!(n-1)!}{
(n+k-1)!(n-s-1)!}$ is a decreasing function of $s$, and $\phi(s)=\frac{(k+s-1)!(n-1)!}{(n+k-1)!(s-1)!}$ is an increasing function $s$. These features are used to obtain Eq.(C42). Eq.(C43) follows from the inequality: $\frac{(k+\lfloor{}n/2\rfloor-1)!(n-1)!}{
(n+k-1)!(\lfloor{}n/2\rfloor-1)!}=
\prod_{i=1}^{\lfloor{}n/2\rfloor}
\frac{(n-i)}{n+k-i}<\frac{n-1}{n+k-1}$.

{\bf Case 3. Proof of Inequalities (C3) and (C4)}

Consider a generalized permutationally symmetric state $|\Phi_{ps}\rangle$ defined in Eq.(12) in the main text. For any state $|\Omega_0\rangle$ in Eq.(C7) we get that
\begin{align*}
|\langle \Omega_0|\Phi_{ps}\rangle|^2
&={\rm tr}[|\Omega_0\rangle\langle\Omega_0|
(\sum_{i,j=0}^{nd-n-1}\alpha_i\alpha_j|D_{i,n}\rangle\langle D_{j,n}|)]
\\
&=\sum_{i=0}^{nd-n-1}\alpha_i^2|\langle \Omega_0|D_{i,n}\rangle|^2
\tag{C46}
\\
&=\left\{
\begin{array}{ll}
D(|\Phi_{ps}\rangle)&\leq \alpha_0^2\beta^2+
\sum_{i=1}^{\lfloor\frac{nd-n-1}{2}\rfloor}\frac{n-1}{n+i-1}(\alpha_i^2+\alpha_{nd-n-1-i}^2),
\\
& \mbox{ for odd } nd-n-1
\\
D(|\Phi_{ps}\rangle)&\leq \alpha_0^2\beta^2+
\sum_{i=1}^{\lfloor\frac{nd-n-1}{2}\rfloor}\frac{n-1}{n+i-1}(\alpha_i^2+\alpha_{nd-n-1-i}^2)
-\frac{2n-2}{nd+n-3}\alpha_{(nd-n-1)/2}^2
\\
& \mbox{ for even } nd-n-1
\end{array}
\right.
\tag{C47}
\end{align*}
where Eq.(C46) follows from the equalities: ${\rm tr}[|\Omega_0\rangle\langle\Omega_0|D_{i,n}\rangle\langle D_{j,n}|]=0$ for $i\not=j$ because $\{|D_{i,n}\rangle\}$ are orthogonal states.  Inequality (C47) follows from Eq.(C2).

{\bf Case 4. Proof of Inequality (C5)}

Consider a genuinely $n$-partite entangled qubit state $|\Phi\rangle$ in the present model. Assume that the Schmidt decomposition of $|\Phi\rangle$ is given by
\begin{align*}
|\Phi\rangle_{A_1\cdots{}A_n}=\sum_{i=0}^{\ell}\lambda_i
|\phi_i\rangle_{A_1\cdots{}A_s}|\psi_i\rangle_{A_{s+1}\cdots{}A_n}
\tag{C48}
\end{align*}
where $\lambda_i\geq0$ and $\sum_{i}\lambda_i^2=1$,
 $|\phi_i\rangle$s are orthogonal states on Hilbert space $\otimes_{i=1}^s\mathbb{H}_{A_i}$, $|\psi_i\rangle$s are orthogonal states on Hilbert space $\otimes_{j=s+1}^n\mathbb{H}_{A_{j}}$. $|\Phi\rangle$ can be regarded as a projection onto the subspace spanned by $\{|\phi_i\rangle|\psi_i\rangle, \forall i\}$. It is sufficient to suppose the decompositions in Eqs.(C31) and (C32), where each qubit Hilbert space $\mathbb{H}_{A_{j}}$ cannot be decomposed into the tensor of two Hilbert spaces with at least two dimensions. Hence, similar to Eqs.(C9) and (C10), we consider that all the decomposed states $|\Psi_1\rangle, \cdots, |\Psi_m\rangle$ in $|\Omega_0\rangle$ in Eq.(C7) are not shared by the same party in order to get $D(|\Phi\rangle)$. From Eqs.(C29)-(C31) we get that
\begin{align*}
|\langle \Phi|\Psi_0\rangle|\Psi_1\rangle|^2 &= (\sum_{i=0}^{\ell}\lambda_ix_iy_i)^2
\\
&\leq \max\{\lambda_0^2, \cdots, \lambda_{\ell}^2\}
\tag{C49}
\end{align*}
from the inequality (C34). Similarly, the inequality (C49) holds for generalized decomposition in Eq.(1) in the main text. From Eq.(C7), it follows that
\begin{align*}
D(|\Phi\rangle)=\max_{{\cal A}_1|{\cal A}_2}\max\{\lambda_0^2, \cdots, \lambda_{\ell}^2\}
\tag{C50}
\end{align*}
where the first maximum is over all the possible bipartitions ${\cal A}_1$ and ${\cal A}_2$ of $\{A_1, \cdots, A_n\}$. This completes the proof of Lemma 3.

\section*{Supplementary D. Proof of Inequality (37)}

It only needs to find the maximal eigenvalues of the reduced density matrices. Consider a general three-qubit pure state as
\begin{align*}
|\Phi\rangle_{ABC}=\lambda_0|000\rangle+\lambda_1e^{i\varphi}|100\rangle+\lambda_2|101\rangle
+\lambda_3|110\rangle+\lambda_4|111\rangle
\tag{D1}
\end{align*}
where $\lambda_i\geq0$, $0\leq\varphi\leq \pi$, and $\sum_{i=0}^4\lambda_i^2=1$.

For the bipartition of $\{A_{1}\}$ and $\{A_2,A_3\}$, we get the reduced density matrix of $A_1$ of $|\Phi\rangle_{ABC}$ as
\begin{align*}
\rho(A_1)=
\left[
\begin{array}{ll}
\lambda_0^2 & \lambda_0\lambda_1
\\
\lambda_0\lambda_1 & 1-\lambda_0^2
\end{array}
\right]
\tag{D2}
\end{align*}
Its maximal eigenvalue is given by
\begin{align*}
\gamma_1=&\frac{1}{2}+
\frac{1}{2}(1-4\lambda_0^2\lambda_2^2
-4\lambda_0^2\lambda_3^2
-4\lambda_0^2\lambda_4^2)^{1/2}
\tag{D3}
\end{align*}

For the bipartition of $\{A_{2}\}$ and $\{A_1,A_3\}$ we get the reduced density matrix of $A_2$ of $|\Phi\rangle_{ABC}$ as
\begin{align*}
\rho(A_2)=
\left[
\begin{array}{ll}
\lambda_0^2+\lambda_1^2+\lambda_2^2
 &
\lambda_1\lambda_3+\lambda_2\lambda_4
\\
\lambda_1\lambda_3+\lambda_2\lambda_4 &
\lambda_3^2+\lambda_4^2
\end{array}
\right]
\tag{D4}
\end{align*}
Its maximal eigenvalue is given by
\begin{align*}
\gamma_2=&\frac{1}{2}+\frac{1}{2}
(1+8\lambda_1\lambda_3\lambda_2\lambda_4
-4\lambda_1^2\lambda_4^2
-4\lambda_2^2\lambda_3^2
-4\lambda_0^2\lambda_4^2
-4\lambda_0^2\lambda_3^2
)^{1/2}
\tag{D5}
\end{align*}

For the bipartition of $\{A_{3}\}$ and $\{A_1,A_2\}$ we get the reduced density matrix of $A_3$ of $|\Phi\rangle_{ABC}$ as
\begin{align*}
\rho(A_{3})=
\left[
\begin{array}{ll}
\lambda_0^2+\lambda_1^2+\lambda_3^2 & \lambda_2\lambda_1+\lambda_3\lambda_4
\\
\lambda_2\lambda_1+\lambda_3\lambda_4 &       \lambda_2^2+\lambda_4^2
\end{array}
\right]
 \tag{D6}
\end{align*}
Its maximal eigenvalue is given by
\begin{align*}
\gamma_3=&\frac{1}{2}+\frac{1}{2}
(1+8\lambda_1\lambda_2\lambda_3\lambda_4
-4\lambda_0^2\lambda_4^2-4\lambda_0^2\lambda_2^2
-4\lambda_1^2\lambda_4^2-4\lambda_2^2\lambda_3^2)^{1/2}
\tag{D7}
\end{align*}
Hence, we get that $\gamma=\max\{\lambda_1,\lambda_2,\lambda_3\}$.

\section*{Supplementary E. Proof of Inequality (44)}

Our goal is to present one method using only two-body correlations in order to witness entangled states in the biseparable model \cite{Sy} inspired by Wigner-Yanase skew information \cite{WY}. Although the goal of the inequality (44) in the main text is to verify multipartite entangled state in the biseparable model \cite{Sy}, we present the bipartite case for the completeness of this method. Interestingly, from Theorem 2, all the genuinely multipartite entangled pure states in the biseparable model \cite{Sy} are also new genuinely multipartite entangled in the present model given in Definition 1. It means that the present inequality (44) provides the first Bell inequality for verifying new genuinely multipartite entangled pure states in the present model given in Definition 1.

\subsection*{E1. Bipartite entangled states}

\begin{figure}
\begin{center}
\resizebox{340pt}{140pt}{\includegraphics{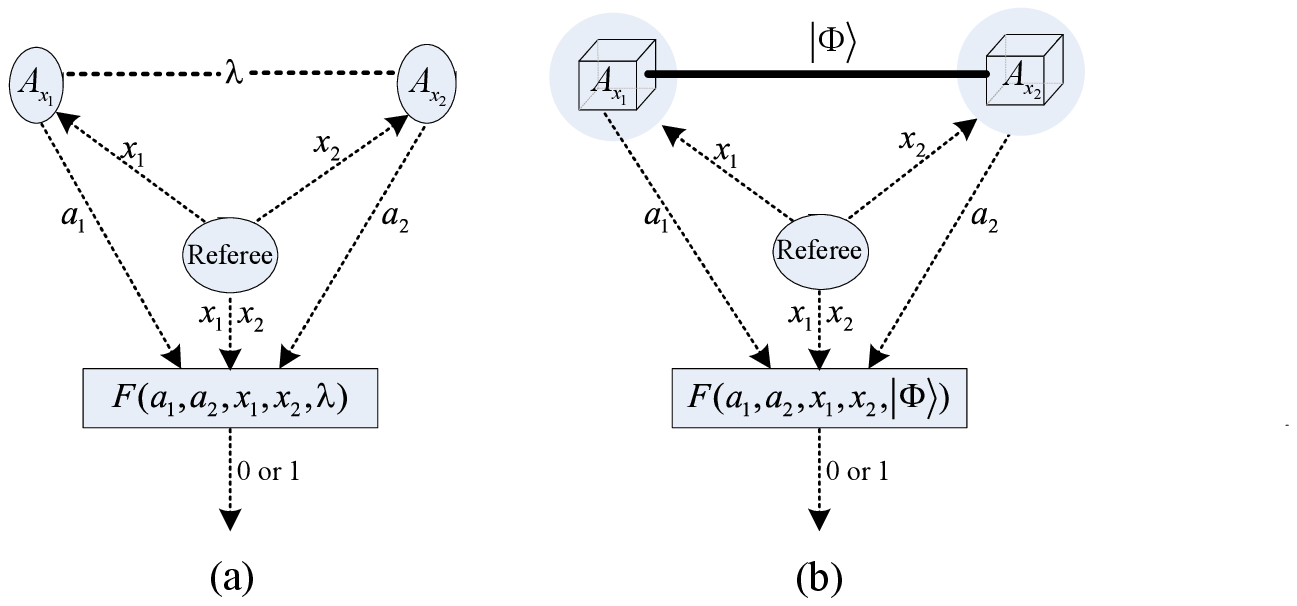}}
\end{center}
\caption{\small (Color online) The Bell test derived from the inequality (E1). Here, some functional box $F$ (depending $f_i$ in Eqs.(E2) and (E3)) that depends on the shared sources will be used. (a) The classical hidden variable model. Two parties share one hidden variable $\lambda$ or a bipartite separable state $\rho=\sum_ip_i\rho^{(1)}_i\otimes\rho^{(2)}_i$, where $\{p_i\}$ is a probability distribution and $\rho^{(j)}_i$s are density operators of the $j$-th system, $j=1, 2$. (b) The quantum entangled system. Two parties share one entangled pure state $|\Phi\rangle$. The measurement operator $A_{x_i}$ depends on the input $x_i$, $i=1,2$. In what follows, we denote Figure x in Supplementary file as Figure Sx for convenience, which is used to distinguish these from the main text.}
\label{fig-S1}
\end{figure}

For a given positive semi-definite operator $\rho$, and a measurement operator $A$, the Wigner-Yanase skew information is defined by $I(A,\rho)=-\frac{1}{2}{\rm tr}[\rho^{p},A]^{1-p}$, where $[A,B]=AB-BA$ denotes the Lie bracket operator. Note that $I(A,\rho)$ is convex for $\rho$ \cite{Lieb}. In what follows, we define similar information with tight upper bound. Actually, we can prove that the following inequality
\begin{align*}
&2\langle{}A\otimes{}B\rangle_{\rho}-\frac{1}{2}f_1(\rho,A)
-\frac{1}{2}f_1(\rho,B)
-f_2(\rho,A,B)\leq 0
\tag{E1}
\end{align*}
for any bipartite separable state $\rho$, and measurement operators $A, B$, where $\langle{}A\otimes{}B\rangle_{\rho}:={\rm tr}[(A\otimes{}B)\rho]$, $f_1$ and $f_2$ are two nonlinear functionals depending on the shared state $\rho$ and measurement operators, which are defined by
\begin{align*}
&f_1(\rho,X)={\rm tr}[\rho^p(X\otimes \mathbbm{1})\rho^{1-p}(X\otimes \mathbbm{1})]
\tag{E2}
\\
&f_2(\rho,A,B)={\rm tr}[\rho^{p}(A\otimes{}\mathbbm{1})
\rho^{1-p}(\mathbbm{1}\otimes{}B)]
\tag{E3}
\end{align*}
with $p\in (0,1)$. Interestingly, the maximum of $2\langle{}A\otimes{}B\rangle-\frac{1}{2}f_1(\rho,A)
-\frac{1}{2}f_1(\rho,B)
-f_2(\rho,A,B)$ for quantum states is $2$. The inequality (E1) can be used to verify bipartite quantum entanglement as shown in Figure S\ref{fig-S1}. There are nonlinear functionals $F_i$s that depend on the shared sources. Fortunately, the classical upper bound is computable in theory.

{\bf Proof of Inequality (E1)}. Note that $-f_i(\rho,X)$ is convex functional for $\rho$ \cite{Lieb}, i.e.,
\begin{align*}
&-f_1(\lambda\rho_1+(1-\lambda)\rho_2,A)\leq -\lambda{}f_1(\rho_1,A)-(1-\lambda)f_1(\rho_2,A)
\\
&-f_2(\lambda\rho_1+(1-\lambda)\rho_2,A,B)\leq -\lambda{}f_2(\rho_1,A,B)-(1-\lambda)f_2(\rho_2,A,B)
\tag{E4}
\end{align*}
where $\rho_1$ and $\rho_2$ are density operators, and $A$ and $B$ are measurement operators. For a given measurement operator $A$, it follows that
\begin{align*}
\max_{\rho=\sum_ip_i\varrho^{(1)}_i\otimes\varrho^{(2)}_i}
\{-f_1(\rho,A)\}
=&\max_{\rho=\varrho^{(1)}\otimes\varrho^{(2)}}\{-f_1(\rho,A)\}
\\
=&\max_{\rho=|\psi_1\rangle\langle\psi_1|\otimes|\psi_2\rangle\langle\psi_2|}\{-f_1(\rho,A)\}
\tag{E5}
\end{align*}
where $\varrho^{(i)}$ can be represented by linearly superposition of pure states, i.e., $\varrho^{(i)}=\sum_jq^{(i)}_j|\psi^{(i)}_j
\rangle\langle\psi^{(i)}_j|$, and $\{q^{(i)}_j\}$ is a probability distribution. Similar result holds for $-f_2(\rho,A,B)$. For a product state $|\psi_1\rangle|\psi_2\rangle$, and two local measurement operators $A$ and $B$, it is easy to check that
\begin{align*}
&f_1(|\psi_1\rangle|\psi_2\rangle,A)=
\langle \psi_1|A|\psi_1\rangle^2,
\\
&f_1(|\psi_1\rangle|\psi_2\rangle,B)=
\langle \psi_2|B|\psi_2\rangle^2,
\\
&f_2(|\psi_1\rangle|\psi_2\rangle,A,B)=
\langle \psi_1|A|\psi_1\rangle\langle \psi_2|B|\psi_2\rangle.
\tag{E6}
\end{align*}
From Eqs.(E5) and (E6), the left side of Eq.(E1) is given
\begin{align*}
\omega:=&\max_{\rho=\sum_ip_i\varrho^{(1)}_i
\otimes\varrho^{(2)}_i}\{2\langle{}A\otimes{}B\rangle_{\rho}
-\frac{1}{2}f_1(\rho,A)
-\frac{1}{2}f_1(\rho,B)-f_2(\rho,A,B)\}
\\
\leq &\max_{\rho=|\psi_1\rangle\langle\psi_1||\psi_2\rangle\langle\psi_2|}
\{2\langle{}A\otimes{}B\rangle_{\rho}
-\frac{1}{2}f_1(\rho,A)
-\frac{1}{2}f_1(\rho,B)-f_2(\rho,A,B)\}
\tag{E7}
\\
=&\max_{\rho=|\psi_1\rangle\langle\psi_1||\psi_2\rangle\langle\psi_2|}
\{\langle{}A\rangle{}_{|\psi_1\rangle}\langle{}B\rangle_{|\psi_2\rangle}
-\frac{1}{2}\langle{}A\rangle_{|\psi_1\rangle}^2
-\frac{1}{2}\langle{}B\rangle^2_{|\psi_2\rangle}\}
\tag{E8}
\\
\leq & 0
\tag{E9}
\end{align*}
which implies the inequality (E1). Eq.(E7) follows from Eq.(E5) and the linearity of the expectation operation $\langle{}\cdot{}\rangle$. Inequality (E8) follows from Eq.(E6). Inequality (E9) follows from the inequality: $(x+y)^2\geq0$.

Note that $-f_1$ and $-f_2$ are convex in the density operator $\rho$ \cite{Lieb}. It only needs to consider pure states for the maximum. The left side of Eq.(E1) is given by
\begin{align*}
\omega_q:=&\max_{\rho=\sum_{i}p_i|\Psi_i\rangle\langle\Psi_i|}
\{2\langle{}A\otimes{}B\rangle_{\rho}
-\frac{1}{2}f_1(\rho,A,B)
-\frac{1}{2}f_1(\rho,A,B)-f_2(\rho,A,B)\}
\\
\leq &\max_{|\Psi\rangle}
\{\langle(A\otimes{}\mathbbm{1}
+\mathbbm{1}\otimes{}B)^2\rangle_{|\Psi\rangle}-\frac{1}{2}f_1(|\Psi\rangle,A)
-\frac{1}{2}f_1(|\Psi\rangle,B)-f_2(|\Psi\rangle,A,B)\}-2
\tag{E10}
\\
=&\max_{|\Psi\rangle}
(\langle(A\otimes{}\mathbbm{1}
+\mathbbm{1}\otimes{}B)^2\rangle_{|\Psi\rangle}
-\frac{1}{2}\langle{}A\otimes{}\mathbbm{1}+\mathbbm{1}\otimes{}B\rangle_{|\Psi\rangle}^2)-2
\tag{E11}
\\
\leq & 2
\tag{E12}
\end{align*}
Eq.(E10) follows from the assumptions of $A, B\leq \mathbbm{1}$. Eq.(E11) follows from the equality:
$f_{1}(|\Psi\rangle, A)=\langle A\otimes\mathbbm{1} \rangle_{|\Psi\rangle}^2$, $f_{1}(|\Psi\rangle, B)=\langle \mathbbm{1}\otimes B \rangle_{|\Psi\rangle}^2$, and $f_{2}(|\Psi\rangle,A,B)=\langle A\otimes B \rangle_{|\Psi\rangle}$. Inequality (E12) makes use of the inequalities: $\langle(A\otimes{}\mathbbm{1}
+\mathbbm{1}\otimes{}B)^2\rangle\leq 4$ and $\langle A\otimes{}\mathbbm{1}+\mathbbm{1}\otimes{}B\rangle^2\geq0$. $\square$

{\bf Example S1}. Consider a bipartite entangled pure state $|\Psi\rangle$ as
\begin{align*}
|\Psi\rangle=\cos\theta|00\rangle+\sin\theta|11\rangle
\tag{E13}
\end{align*}
where $\theta\in(0,\frac{\pi}{2}]$. Define $A=B=\sigma_x$ as Pauli operator. It is easy to evaluate that
\begin{align*}
\omega_q=&2\langle{}A\otimes{}B\rangle_{|\Psi\rangle}-\frac{1}{2}f_1(|\Psi\rangle,A)
-\frac{1}{2}f_1(|\Psi\rangle,B)-\frac{1}{2}f_2(|\Psi\rangle,A,B)]
\\
=&2{\rm tr}[(A\otimes{}B)|\Psi\rangle\langle\Psi|]-
\frac{1}{2}({\rm tr}[(A\otimes{}\mathbbm{1})|\Psi\rangle\langle\Psi|]^2
+{\rm tr}[(\mathbbm{1}\otimes{}B)|\Psi\rangle\langle\Psi|])^2
\\
=&4\sin2\theta
\\
>&0
\tag{E14}
\end{align*}
when $\theta\in (0,\frac{\pi}{2}]$, where ${\rm tr}[(A\otimes{}\mathbbm{1})|\Psi\rangle\langle\Psi|]={\rm tr}[(\mathbbm{1}\otimes{}B)|\Psi\rangle\langle\Psi|]=0$. If define $A=B=\sigma_z$, we get $\omega_q=2-2\cos2\theta^2>0$ for $\theta\in (0,\frac{\pi}{2})$, where ${\rm tr}[(A\otimes{}\mathbbm{1})|\Psi\rangle\langle\Psi|]={\rm tr}[(\mathbbm{1}\otimes{}B)|\Psi\rangle\langle\Psi|]=\cos2\theta$. Hence, two pauli measurement operators are useful to  verify generalized bipartite entangled pure states. Interestingly, the maximally entangled $|\Psi\rangle$ achieves the maximal violation.

Similarly, consider a generalized bipartite entangled pure state on Hilbert space $\mathbb{H}_A\otimes \mathbb{H}_B$ as
\begin{align*}
|\Psi\rangle=\sum_{i=0}^{d-1}a_i|ii\rangle
\tag{E15}
\end{align*}
where $\mathbb{H}_A$ and $\mathbb{H}_B$ have the same dimension $d\geq2$, and $\gamma_i$ satisfies $\sum_{i=0}^{d-1}a_i^2=1$. It is entangled if $a_ia_j\not=0$ for some $i\not=j$. Consider the subspace spanned by $\{|i\rangle, |j\rangle\}$. It is easy to prove that Pauli operators $\sigma_x, \sigma_z$ on the subspace can also be used to verify the nonlocality of $|\Psi\rangle$.

\subsection*{E2. Multipartite entangled states}

Consider an $n$-partite state $\rho$ on Hilbert space $\mathbb{H}_{A_1}\otimes\cdots\otimes \mathbb{H}_{A_n}$ with the same dimension $d\geq 2$. $\rho$ is fully separable \cite{Sy} if the following decomposition holds
\begin{align*}
\rho=\sum_{i}p_i\rho^{(1)}_i\otimes\cdots{}\otimes\rho^{(n)}_i
\tag{E16}
\end{align*}
where $\{p_i\}$ is a probability distribution, and $\rho^{(j)}_i$s are density operators on the local system of $A_j$, $j=1, \cdots, n$. Inequality (E1) can be then extended as:
\begin{align*}
&\sum_{1\leq i\not=j\leq n}\langle{}M_i\otimes{}M_j\rangle_{\rho}-
\frac{n-1}{n}\sum_{i=1}^n(f_1(\rho,M_i)
+\sum_{j\not=i}^nf_2(\rho,M_i,M_j))
\leq 0
\tag{E17}
\end{align*}
for any fully separable state $\rho$, where $f_1$ and $f_2$ are two nonlinear functionals defined in the respective Eqs.(E2) and (E3), and $M_i$s are dichotomic measurement operators. On the other hand, the maximum of l.h.s of Eq.(E17) is $n^2-n$ for general quantum states. This inequality can be used to verify multipartite entangled states.

{\bf Proof of Inequality (E17)}. Similar to bipartite states, $-f_i$s are convex for the density operators with any measurement operators $M_i$ \cite{Lieb}. Consider any fully separable pure state of $\rho_{fs}=\otimes_{i=1}^n|\psi_i\rangle\langle \psi_i|$. The left side of the inequality (E17) is rewritten by:
\begin{align*}
\omega:=&\max_{\rho_{fs}}\{\sum_{1\leq i\not=j\leq n}
\langle{}M_i\otimes{}M_j\rangle_{\rho}-
\frac{n-1}{n}\sum_{i=1}^n(f_1(\rho,M_i)
+\sum_{j\not=i}f_2(\rho,M_i,M_j))\}
\\
\leq&\max_{\rho=\otimes_{i=1}^n|\psi_i\rangle}
\{\sum_{1\leq i\not=j\leq n}\langle{}M_i\otimes{}M_j\rangle_{\rho}-
\frac{n-1}{n}\sum_{i=1}^n(f_1(\rho,M_i)
+\sum_{j\not=i}f_2(\rho,M_i,M_j))\}
\tag{E18}
\\
\leq&\max_{\rho=\otimes_{i=1}^n|\psi_i\rangle}\{\langle (\sum_{j=1}^nM_i)^2\rangle_{\rho} -n^2-\frac{n-1}{n}\sum_{i=1}^n(f_1(\rho,M_i)
-\sum_{j\not=i}f_2(\rho,M_i,M_j))\}
\tag{E19}
\\
=&\max_{\otimes_{i=1}^n|\psi_i\rangle}\{\frac{1}{n}
\sum_{1\leq i\not=j\leq n}\langle{}M_i\rangle_{|\psi_i\rangle}\langle{}M_j\rangle_{|\psi_j\rangle}
-\frac{n-1}{n}\sum_{i=1}^n\langle{}M_i\rangle_{|\psi_i\rangle}^2\}-n
\tag{E20}
\\
\leq &0
\tag{E21}
\end{align*}
Inequality (E18) follows from the convexity of $-f_i$s. Inequality (E19) follows from $M_i\leq \mathbbm{1}$. Eq.(E20) follows from the equalities: $f_1(\otimes_{i=1}^n|\psi_i\rangle,M_j)=(\langle {}\psi_i|\otimes_{i=1}^n)M_j(\otimes_{k=1}^n|\psi_k\rangle):=\langle{}M_j\rangle_{|\psi_j\rangle}$. Inequality (E21) follows from the inequality: $\sum_{1\leq i\not=j\leq n}\langle{}M_i\rangle_{|\psi_i\rangle}\langle{}M_j\rangle_{|\psi_j\rangle}
-(n-1)\sum_{i=1}^n\langle{}M_i\rangle_{|\psi_i\rangle}^2=\sum_{1\leq i< j\leq n}(\langle{}M_i\rangle_{|\psi_i\rangle}+\langle{}M_j\rangle_{|\psi_j\rangle})^2\geq0$.

For an $n$-partite quantum state of $\rho=\sum_ip_i|\Phi_i\rangle\langle\Phi_i|$, with  similar evaluation, the left side of Eq.(E17) is rewritten by:
\begin{align*}
\omega_{q}:=&\max_{\rho}\{\sum_{1\leq i\not=j\leq n}\langle{}M_i\otimes{}M_j\rangle_{\rho}-
\frac{n-1}{n}\sum_{i=1}^n(f_1(\rho,M_i)
+\sum_{j\not=i}f_2(\rho,M_i,M_j))\}
\\
\leq &\max_{\rho=|\Phi\rangle\langle\Phi|}\{\sum_{1\leq i\not=j\leq n}
\langle{}M_i\otimes{}M_j\rangle_{\rho}-
\frac{n-1}{n}\sum_{i=1}^n(f_1(\rho,M_i)
+\sum_{j\not=i}f_2(\rho,M_i,M_j))\}
\tag{E22}
\\
\leq &\max_{|\Phi\rangle}
\{\langle{}(\sum_{i=1}^nM_i)^2\rangle_{|\Phi\rangle}-
\frac{n-1}{n}\sum_{i=1}^n(f_1(|\Phi\rangle,M_i)
+\sum_{j\not=i}f_2(|\Phi\rangle,M_i,M_j))\}-n
\tag{E23}
\\
=&\max_{|\Phi\rangle}
\{\langle{}(\sum_{i=1}^nM_i)^2\rangle_{|\Phi\rangle}-
\frac{n-1}{n}(\sum_{i=1}^n\langle M_i\rangle_{|\Phi\rangle})^2\}-n
\tag{E24}
\\
\leq &n^2-n
\tag{E25}
\end{align*}
Inequality (E22) follows from the convexity of $-f_i$s. Inequality (E23) follows from the assumptions of $M_i\leq \mathbbm{1}$ for all $M_i$s. Eq.(E23) follows from the equalities: $f_1(|\Phi\rangle,M_j)=\langle {}\Phi|M_j|\Phi\rangle:=\langle{}M_j\rangle$. Inequality (A24) follows from the inequalities: $\langle{}(\sum_{i=1}^nM_i)^2\rangle\leq n^2$ and $(\sum_{i=1}^n\langle M_i\rangle_{|\Phi\rangle})^2\geq0$. $\square$

In what follows, some examples will be presented to show the maximal violation of the inequality (E17).

{\bf Example S2}. Consider an $n$-partite generalized GHZ state \cite{GHZ}:
\begin{align*}
|\Phi\rangle=\cos\theta|0\rangle^{\otimes n}+\sin\theta|1\rangle^{\otimes n}
\tag{E26}
\end{align*}
with $\theta\in (0,\frac{\pi}{4}]$. Define $M_i=\sigma_z$, $i=1, \cdots, n$. It follows that
\begin{align*}
\omega_q(|\Phi\rangle)=&n^2-n-n(n-1)\cos2\theta^2
>0
\tag{E27}
\end{align*}
for any $\theta\in (0,\frac{\pi}{4}]$. It means that generalized GHZ states are not fully separable for any $\theta\in (0,\frac{\pi}{4}]$.

{\bf Example S3}. Consider an $n$-partite generalized W state
\begin{align*}
|W\rangle=\sum_{i=1}^n\alpha_i|0\rangle^{\otimes i-1}|1\rangle|0\rangle^{\otimes n-i}
\end{align*}
with $\sum_{i=1}^n\alpha_i^2=1$ and $\alpha_i\geq0$. Define $M_i=\sigma_x$, $i=1, \cdots, n$. It follows that
\begin{align*}
\omega_q(|W\rangle)=&\sum_{1\leq i\not=j\leq n}4\alpha_i\alpha_j
\\
> & 0
\tag{E28}
\end{align*}
for $\alpha_i\alpha_j\not=0$ with any $i\not=j$, which implies a generalized singlet state for subsystems $i,j$.

{\bf Example S4}. Consider a generalized $4$-partite Dicke state \cite{Dicke1}:
\begin{align*}
|D_{2,4}\rangle=&\gamma_1|0011\rangle+\gamma_2|1100\rangle
+\gamma_3|0101\rangle
+\gamma_4|1010\rangle+\gamma_5|1001\rangle+\gamma_6|0110\rangle
\tag{S29}
\end{align*}
where $\gamma_i$s satisfy $\sum_{i=1}^6\gamma_i^2=1$. Define $M_i=\sigma_x, i=1, \cdots, n$. It follows that
\begin{align*}
\omega_q(|D_{2,4}\rangle)=&\sum_{1\leq i<j\leq 6} 8\gamma_i\gamma_j
>0
\tag{S30}
\end{align*}
for all nonnegative $\gamma_i$ with at least two $\gamma_i,\gamma_j\not=0$.

Now, we consider genuinely multipartite entangled states in the biseparable model. For a general $n$-partite state $\rho$ on Hilbert space $\mathbb{H}_{A_1}\otimes{}\cdots{}\otimes\mathbb{H}_{A_n}$, it is $k$-producible state \cite{Sy} if the following decomposition holds
\begin{align*}
\rho_{k\mbox{\small-pr}}=\sum_{I_1, \cdots, I_k}\sum_ip_i\rho^{(I_1)}_i\otimes{}\cdots\otimes \rho^{(I_k)}_i
\tag{E31}
\end{align*}
where $\{I_1, \cdots, I_k\}$ is a $k$-partite partition of $\{1, \cdots, n\}$, i.e., $\cup_{i=1}^kI_i=\{1, \cdots, n\}$ and $I_i\cap{}I_j=\emptyset$ for $i\not=j$, $\rho^{(I_j)}_i$ are density operators of local system $I_j$. We can prove the following inequality
\begin{align*}
&\sum_{1\leq i\not=j\leq n}\frac{1}{n}\langle{}M_i\otimes{}M_j\rangle_{\rho}-
\frac{n-1}{n^2}\sum_{i=1}^n(f_1(\rho,M_i)
+\sum_{j\not=i}^nf_2(\rho,M_i,M_j))\leq
 k-1
\tag{E32}
\end{align*}
for all $k$-producible $n$-partite states $\rho$. The upper bound is $n-1$ for any biseparable states \cite{Sy}.

{\bf Proof of Inequality (E32)}. Actually, the left side of Eq.(E32) is rewritten by:
\begin{align*}
\omega_{k}:=&\max_{\rho_{k\mbox{\small-pr}}}
\sum_{1\leq i\not=j\leq n}
\frac{1}{n}\langle{}M_i\otimes{}M_j\rangle_{\rho_{k\mbox{\small-pr}}}
-\frac{n-1}{n^2}\sum_{i=1}^n(f_1(\rho_{\rho_{k\mbox{\small-pr}}},M_i)
+\sum_{j\not=i}f_2(\rho_{\rho_{k\mbox{\small-pr}}},M_i,M_j))\}
\\
\leq &\max_{\rho=\otimes_{i=1}^m|\Phi_i\rangle_{I_i}\langle\Phi_i|}\{\sum_{1\leq i\not=j\leq n}
\frac{1}{n}\langle{}M_i\otimes{}M_j\rangle_{\rho}
-\frac{n-1}{n^2}\sum_{i=1}^n(f_1(\rho,M_i)
+\sum_{j\not=i}f_2(\rho,M_i,M_j))\}
\tag{E33}
\\
\leq&\max_{\rho=\otimes_{i=1}^m|\Phi_i\rangle_{I_i}\langle\Phi_i|}\{
\frac{1}{n}\langle{}(\sum_{i=1}^nM_i)^2\rangle_{\rho}
-\frac{n-1}{n^2}\sum_{i=1}^n(f_1(\rho,M_i)
+\sum_{j\not=i}f_2(\rho,M_i,M_j))\}-1
\tag{E34}
\\
=&\frac{1}{n^2}\max_{\rho=\otimes_{i=1}^m|\Phi_i\rangle_{I_i}\langle\Phi_i|}
\{\langle{}(\sum_{i=1}^nM_i)^2\rangle_{\rho}+(n-1)
I(\rho,\sum_{i=1}^nM_i)\}-1
\tag{E35}
\\
\leq&\frac{1}{n^2}(nk+(n-1)\max_{n_1+\cdots+n_m=n,
\atop{1\leq n_1, \cdots, n_m\leq k}}\sum_{i=1}^mn_i^2)-1
\tag{E36}
\\
\leq &\frac{1}{n}\times{}k^2\times \frac{n}{k}-1
\tag{E37}
\\
=&k-1
\tag{E38}
\end{align*}
Inequality (E33) follows from the convexity of $-f_i$s \cite{Lieb}. Inequality (E34) follows from the assumptions of $M_i\leq \mathbbm{1}$ for all $i$s. Eq.(E35) follows from the definition of Wigner-Yanase skew information $I$ \cite{WY}. Note that $I(|\phi_i\rangle|\phi_j\rangle, M_i+M_j)
=I(|\phi_i\rangle,M_i)+I(|\phi_j\rangle,M_j)$ holds for any product states. From the inequality of $\max_{\rho=\otimes_{i=1}^m|\Phi_i\rangle_{I_i}\langle\Phi_i|}\langle{}
(\sum_{i=1}^nM_i)^2\rangle_{\rho}< \max_{\rho=\otimes_{i=1}^m|\Psi_i\rangle_{I_i}\langle\Phi_i|}(n-1)
I(\rho,\sum_{i=1}^nM_i)$, it follows that
$\max_{\rho=\otimes_{i=1}^m|\Phi_i\rangle_{I_i}\langle\Phi_i|}
(\langle(\sum_{i=1}^nM_i)^2\rangle_{\rho}
+(n-1)I(\rho,\sum_{i=1}^nM_i))$ given in the right side of Eq.(E36) is achieved for $\rho=\otimes_{i=1}^m|\Phi_i\rangle_{I_i}\langle\Phi_i|$, which maximizes $(n-1)I(\rho,\sum_{i=1}^nM_i)$. In this case, $\langle(\sum_{i=1}^nM_i)^2\rangle_{\rho}=\sum_{j=1}^m\langle{}
(\sum_{I_j}M_s)^2\rangle_{\rho}
\leq k^2\times \frac{n}{k}=nk$. This has proved the inequality (E37). Inequality (E38) has taken use of the following inequality: $\sum_{i=1}^mn_i^2\leq (m-1)k^2+n_m^2\leq k^2\times \frac{n}{k}$, where $n_1=n_2=\cdots=n_{m-1}=k$.

The maximum of the quantity in the left side of Eq.(E32) is $n-1$ for $n$-partite genuinely entangled state in the biseparable model \cite{Sy}. $\square$

This witness is device-independent in the sense that any observed violation of $\omega_{k-pr}$ by $\rho$ implies that $\rho$ is at least genuinely $k+1$-partite entangled, i.e., it has an entanglement depth of at least $k+1$ in the biseparable model \cite{Sy}, regardless of the details of measurement devices and Hilbert space dimensions.

\begin{figure}
\begin{center}
\resizebox{200pt}{160pt}{\includegraphics{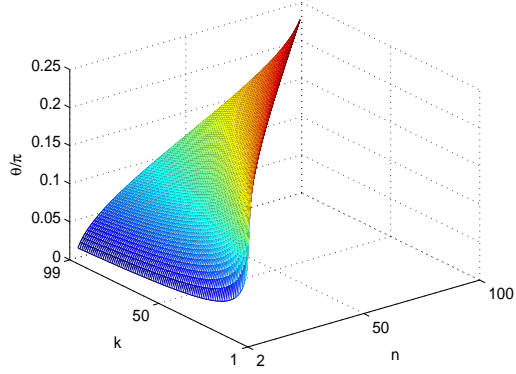}}
\end{center}
\caption{\small (Color online) The lower bound of $\theta$ for verifying the entanglement depth of generalized GHZ states in terms of the number $n$ of particles in the biseparable model \cite{Sy}.}
\label{fig-s2}
\end{figure}

{\bf Example S5}. Consider an $n$-partite generalized GHZ states $|\Phi\rangle=\cos\theta|0\rangle^{\otimes n}+\sin\theta|1\rangle^{\otimes n}$ with $\theta\in (0,\frac{\pi}{2}]$. Define $M_i=\sigma_z$, $\gamma_i=1$ for all $i$s. It follows that
\begin{align*}
\omega_{k\mbox{\small -pr}}(|\Phi\rangle)=&n-1-(n-1)\cos2\theta^2
\\
>&k-1
\tag{E39}
\end{align*}
when $\cos2\theta<\sqrt{\frac{n-k}{n-1}}$. It implies that the inequality (E32) can be used to verify the entanglement depth $k\geq 2$ for the maximally GHZ state or partially entangled GHZ states. Some numerical simulations of $\theta$ are shown in Figure S\ref{fig-s2}. Specially, when $k=n-1$, it shows that the inequality (E32) can be used to verify the genuinely multipartite nonlocality of partially entangled GHZ states with $\cos2\theta<\sqrt{\frac{n-2}{n-1}}$ in the biseparable model \cite{Sy}. From Theorem 2, it means that the inequality (E32) can be used to verify new genuinely multipartite nonlocality of partially entangled GHZ states with $\cos2\theta<\sqrt{\frac{n-2}{n-1}}$ in the present model.

{\bf Example S6}. Consider an $n$-partite W state $|W_n\rangle=\sum_{i=1}^n\gamma_i|0\rangle^{\otimes i-1}|1\rangle|0\rangle^{\otimes n-i}$ with $\sum_i\gamma_i^2=1$. Define $A_i=\sigma_x$ for all $i$s. It follows that
\begin{align*}
\omega_{k\mbox{\small -pr}}(|W_n\rangle)=&\frac{\sum_{i\not=j}4\gamma_i\gamma_j}{n}
\\
>&1
\tag{E40}
\end{align*}
when $\gamma_i=\frac{1}{\sqrt{n}}$ and $k=2$. It implies that the inequality (E32) can be used to verify the entanglement depth $k=2$ of the maximally entangled W states in the biseparable model \cite{Sy}.

\begin{figure}
\begin{center}
\resizebox{200pt}{160pt}{\includegraphics{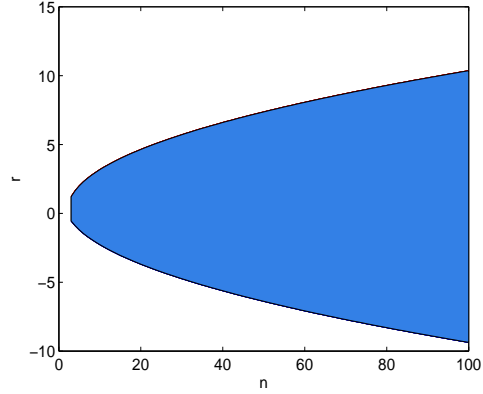}}
\end{center}
\caption{\small (Color online) The possible values of $r$ for verifying the entanglement depth $k=2$ of generalized W states in terms of the number $n$ of particles in the biseparable model \cite{Sy}.}
\label{fig-s3}
\end{figure}

The second example is a generalized $n$-particle W state
\begin{align*}
|W_n\rangle=&\frac{1}{\sqrt{n-1+r^2}}\sum_{i=1}^{n-1}|0\rangle^{\otimes i-1}|1\rangle|0\rangle^{\otimes n-i}
+\frac{r}{\sqrt{n-1+r^2}}|0\rangle^{\otimes n-1}|1\rangle
\tag{E41}
\end{align*}
The maximum of $\omega_{2\mbox{\small -pr}}(|W_n\rangle)$ with $k=2$ is given by
\begin{align*}
\omega_{2\mbox{\small -pr}}(|W_n\rangle)=&\frac{2(n-1)^2+r(n-1)}{n(n-1+r^2)}
\\
>&1
\tag{E42}
\end{align*}
when $r$ satisfies $\frac{1}{2n}(n-1-2\sqrt{(n-1)(n^2+2n-4)})
<r<\frac{1}{2n}(n-1+2\sqrt{(n-1)(n^2+2n-4)})$, as shown in Figure S\ref{fig-s3}.

Another example is a generalized $n$-particle W state defined by
\begin{align*}
|W_n\rangle=&\frac{r}{\sqrt{1+(n-1)r^2}}\sum_{i=1}^{n-1}|0\rangle^{\otimes i-1}|1\rangle|0\rangle^{\otimes n-i}
+\frac{1}{\sqrt{1+(n-1)r^2}}|0\rangle^{\otimes n-1}|1\rangle
\tag{E43}
\end{align*}
We get that
\begin{align*}
\omega_{2\mbox{\small -pr}}(|W_n\rangle)=&\frac{8r+4r^2}{1+2r^2}
>1
\tag{E44}
\end{align*}
for any $r>\frac{-2(n-1)+\sqrt{(n-1)(n^2+2n-4)}}{(n-1)(n-2)}$. It implies that the inequality (E32) can be used to verify the entanglement depth $k=2$ of generalized W states.

\begin{figure}
\begin{center}
\resizebox{260pt}{240pt}{\includegraphics{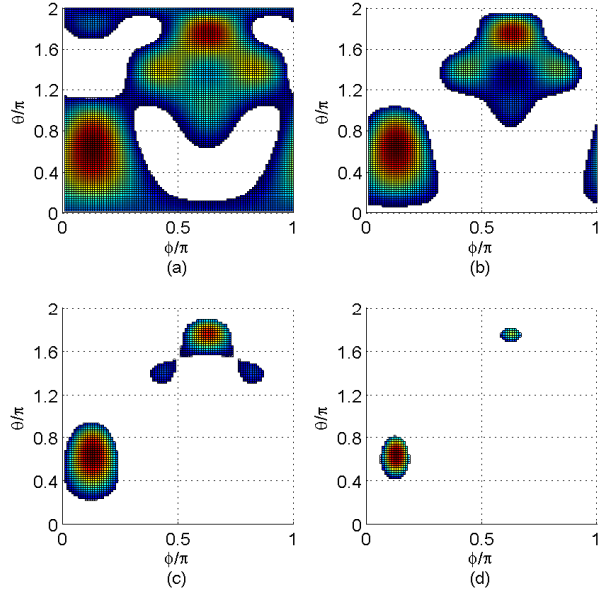}}
\end{center}
\caption{\small (Color online) The values of $\phi,\theta$ for verifying the entanglement depth $k$ of generalized Dicke state $|D_{2,4}\rangle$. (a) $k=1$ for fully separable states. (b) $k=2$. (c) $k=3$. (d) $k=4$.}
\label{fig-s4}
\end{figure}

{\bf Example S7}. Consider a generalized 4-partite Dicke state \cite{Dicke1} shown in Eq.(A29). Define $A_i=\sigma_x$ for all $i$. It follows that
\begin{align*}
\omega_{k\mbox{\small -pr}}(|D_{2,4}\rangle)=&\frac{4}{3}\sum_{1\leq i<j\leq 6}\gamma_i\gamma_j
\tag{E45}
\end{align*}
Consider $\gamma_1=\gamma_2=\frac{1}{\sqrt{2}}\sin\phi\cos\theta$, $\gamma_3=\gamma_4=\frac{1}{\sqrt{2}}\sin\phi\sin\theta$ and $\gamma_5=\gamma_6=\frac{1}{\sqrt{2}}\cos\phi$, where $\phi,\theta$ are phases of sphere coordinates, i.e., $\phi\in[0,\pi]$ and $\theta=[0,2\pi]$. It follows from Eq.(E45) that
\begin{align*}
\omega_{k\mbox{\small -pr}}(|D_{2,4}\rangle)=&
\frac{4}{3}\sin\phi^2\sin2\theta+\frac{4}{3}\sin2\phi(\cos\theta+\sin\theta)
+\frac{2}{3}
\\
>&k-1
\tag{E46}
\end{align*}
for some $\phi,\theta$ shown in Figure S\ref{fig-s4}. It implies that the inequality (E32) can be used to verify all the entanglement depths $k$ of some generalized Dicke states. Interestingly, it can be used to verify the genuinely multipartite nonlocality of Dicke states in the biseparable model \cite{Sy} going beyond the maximally entangled Dicke state with $\gamma_i=\frac{1}{\sqrt{6}}$, i.e., $\omega_{k\mbox{\small -pr}}(|D_{2,4}\rangle)>3$ for some $\phi,\theta$. From Theorem 2, it provides a method verify new genuinely multipartite nonlocality of Dicke states in the present model given in Eq.(1) in the main text.

So far, all the examples are permutationally symmetric multipartite states. It is also useful for verifying asymmetric states.

\begin{figure}
\begin{center}
\resizebox{260pt}{240pt}{\includegraphics{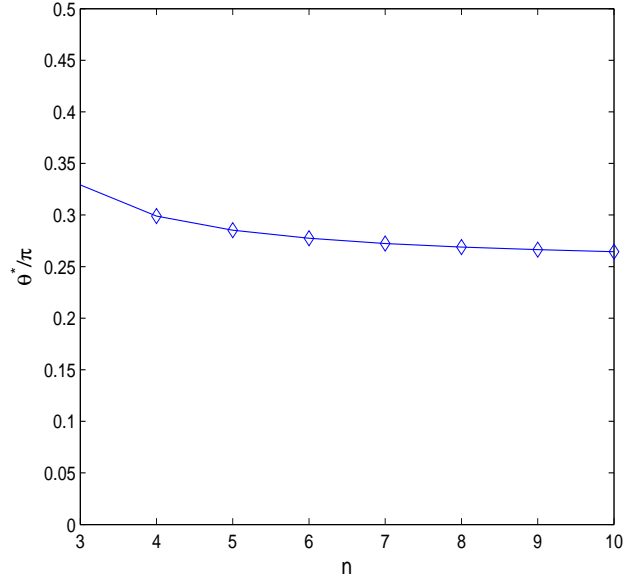}}
\end{center}
\caption{\small (Color online) The possible values $\theta$ with $\theta<\theta^*$ for verifying the genuinely $n$-partite nonlocality of $n$-partite maximal slice (MS) state.}
\label{fig-s5}
\end{figure}

{\bf Example S8}. Consider an $n$-partite maximal slice (MS) state \cite{MS}:
\begin{align*}
|\Phi_s\rangle=\frac{1}{\sqrt{2}}(|0\rangle^{\otimes n}+|1\rangle^{\otimes n-1}(\cos\theta|0\rangle+\sin\theta|1\rangle)
\tag{E47}
\end{align*}
with $\theta\in(0,\frac{\pi}{2}]$. Define $A_i=\sigma_z$ for all $i$s. It follows that
\begin{align*}
\omega(|\Phi_s\rangle)=&
\frac{(n-1)(n-2)}{n}+\frac{n-1}{n}(1+\cos2\theta)
-\frac{n-1}{4n^2}(1+\cos2\theta)^2
\\
>& n-2
\tag{E48}
\end{align*}
when $\theta$ satisfies the inequality: $\cos2\theta>\frac{2n^2-3n+1-2\sqrt{n^4-3n^3+4n^2-2n}}{n-1}$. It means that $n$-partite maximal slice (MS) state has the genuinely $n$-partite nonlocality in the biseparable model \cite{Sy}. From Theorem 2, it provides a method verify new genuinely multipartite nonlocality of MS states in the present model given in Eq.(1) in the main text. The numeric evaluations are shown in Figure S\ref{fig-s5}.

{\bf Example S9}. Consider a generalized tripartite entangled pure state \cite{AAC}:
\begin{align*}
|\Phi_3\rangle=\gamma_0|000\rangle
+\gamma_1e^{i\phi}|100\rangle
+\gamma_2|101\rangle+\gamma_3|110\rangle+\gamma_4|111\rangle
\tag{E49}
\end{align*}
where $\phi\in[0,\pi],\gamma_i\geq0,
\sum_{i}\gamma_i^2=1,\gamma_0\not=0,\gamma_2+\gamma_4\not=0,\gamma_3+\gamma_4\not=0$.
Define $A_i=\sigma_z$ for all $i$s. It follows that
\begin{align*}
\omega_{2\mbox{\small-pr}}(|\Phi_3\rangle)=&2\gamma_0^2+2\gamma_4^2-
\frac{2}{3}(\gamma_1^2+\gamma_2^2+\gamma_3^2)
-
\frac{4}{9}(3\gamma_0^2+\gamma_1^2-\gamma_2^2-\gamma_3^2-\gamma_4^2)^2
\\
>& 1
\tag{E50}
\end{align*}
which violates the inequality (E32) when $\gamma_i$ satisfy the inequality: $\gamma_0^2+\gamma_4^2>\frac{1}{2}$ and  $3\gamma_0^2+\gamma_1^2=\gamma_2^2+\gamma_3^2+\gamma_4^2$. i.e., $\gamma_0^2=\frac{1}{6}-\frac{1}{3}\gamma_1^2, \gamma_4^2=\frac{1}{2}-\gamma_2^2-\gamma_3^2$ and $\gamma_2^2+\gamma_3^2+\frac{1}{3}\gamma_1^2\leq \frac{1}{6}$.

\subsection*{E3. Mixed entangled states}

Consider the $n$-partite Werner state \cite{Werner}:
\begin{align*}
\rho=v_n|\Phi\rangle\langle \Phi|+\frac{1-v_n}{2^n}\mathbbm{1}_{2^n}
\tag{E51}
\end{align*}
where $|\Phi\rangle$ is any $n$-partite qubit state on Hilbert space $\mathbb{H}_{1}\otimes \cdots\otimes\mathbb{H}_n$, and $\mathbbm{1}_{2^n}$ is the identity operator, $v_n\in[0,1]$ is a noise parameter. For a positive semidefinite operator of $\rho$ we have
\begin{align*}
\rho^{1/2}=&g_v|\Phi\rangle\langle \Phi|+\sqrt{\frac{1-v_n}{2^n}}\mathbbm{1}_{2^n}
\tag{E52}
\end{align*}
with  $g_v=\sqrt{v_n+\frac{1-v_n}{2^n}}-\sqrt{\frac{1-v_n}{2^n}}$. The nonlinear functionals $f_i$s can be then rewritten into
\begin{align*}
&f_1(\rho,A_i)=g_v^2\langle\Phi|A_i|\Phi\rangle^2
+2g_v\sqrt{\frac{1-v_n}{2^n}}\langle\Phi|A_i|\Phi\rangle
-v_n+1
\tag{E53}
\\
&f_2(\rho,A_i,A_j)=g_v^2\langle\Phi|A_i|\Phi\rangle\langle\Phi|A_j|\Phi\rangle
+2g_v\sqrt{\frac{1-v}{2^n}}\langle\Phi|A_i\otimes{}A_j|
\Phi\rangle
\tag{E54}
\end{align*}
for $p=\frac{1}{2}$. It follows from Eqs.(E53), (E54) and (E32) that
\begin{align*}
&\sum_{1\leq i\not=j\leq n}(\frac{1}{n}-\frac{2(n-1)g_v}{n^2v_n}
\sqrt{\frac{1-v_n}{2^n}})\langle{}A_iA_j\rangle_{\rho}-
\frac{n-1}{n^2}\frac{g_v^2}{v_n^2}
(\sum_{i=1}^n\langle{}A_i\rangle_{\rho})^2
\\
&
-\frac{2(n-1)g_v}{n^2v_n}\sqrt{\frac{1-v_n}{2^n}}
\langle{}A_i\rangle_{\rho}
-\frac{(n_n-1)(1-v_n)}{n}
\leq k-1
\tag{E55}
\end{align*}
for any $k$-producible state $\rho$. Note that the noise parameter is involved in the left side of the inequality (E55). Hence, it should be computed firstly in applications.

\begin{figure}
\begin{center}
\resizebox{260pt}{200pt}{\includegraphics{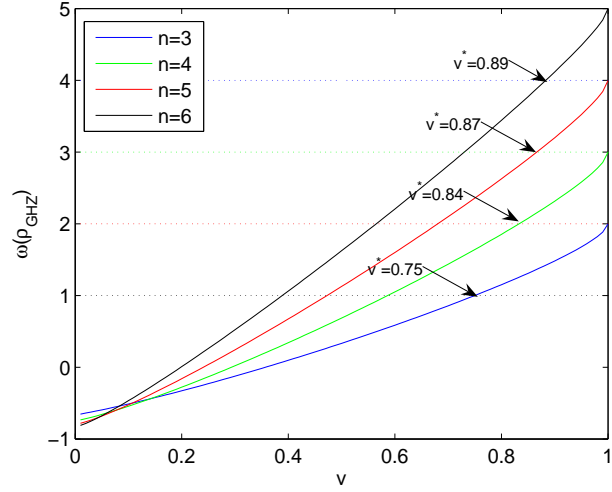}}
\end{center}
\caption{\small (Color online) $v^*_n$ denotes the  critical noise parameter $v$ for verifying the genuinely $n$-partite nonlocality of Werner state $\rho_{GHZ}$ in the biseparable model \cite{Sy}, $n=3, \cdots, 6$.}
\label{fig-s6}
\end{figure}

{\bf Example S10}. Consider an $n$-partite GHZ state: $|\Phi\rangle=\frac{1}{\sqrt{2}}(|0\rangle^{\otimes n}+|1\rangle^{\otimes n})$. Define $A_i=\sigma_z$ for all $i$. It follows that $\langle{}A_i\rangle_{\rho_{GHZ}}=0$, and  $\langle{}A_i\otimes{}A_j\rangle_{\rho_{GHZ}}=v_n$ for all $i$ and $j\not=i$. From Eqs.(E51) and (E55) we get that
\begin{align*}
\omega_{k\mbox{\small-pr}}(\rho_{GHZ})=&\sum_{i\not=j}(\frac{1}{n}
-\frac{2(n-1)g_v}{n^2v_n}
\sqrt{\frac{1-v_n}{2^n}})\langle{}A_i\otimes{}A_j\rangle_{\rho_{GHZ}}
-\frac{n-1}{n^2}\frac{g_v^2}{v_n^2}(\sum_{i=1}^n\langle{}A_i\rangle_{\rho_{GHZ}})^2
\\
&-\frac{2(n-1)g_v}{n^2v_n}\sqrt{\frac{1-v_n}{2^n}}\langle{}A_i\rangle_{\rho_{GHZ}}
-\frac{(n-1)(1-v_n)}{n}
\\
=&v_n(n-1)-\frac{2(n-1)^2g_v}{n}\sqrt{\frac{1-v_n}{2^n}}
-\frac{(n-1)(1-v_n)}{n}
\\
> &k-1
\tag{E56}
\end{align*}
which violates the inequality (E55) when $v_n> v^*_n$. Numeric evaluations of $v$ are shown in Figure S\ref{fig-s6}.

\begin{figure}
\begin{center}
\resizebox{260pt}{200pt}{\includegraphics{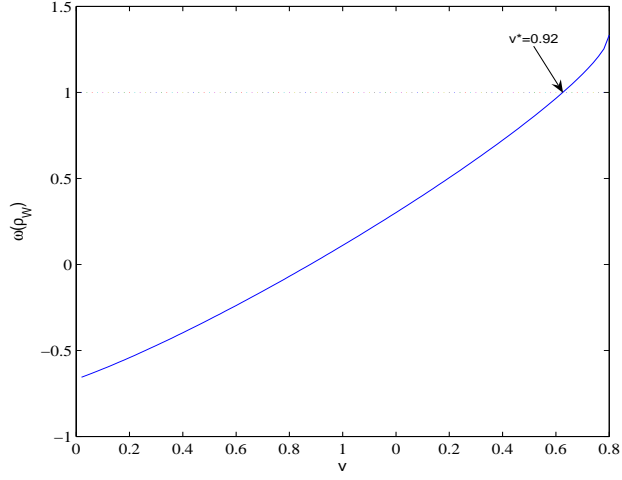}}
\end{center}
\caption{\small (Color online) $v^*$ denotes the  critical noise parameter $v$ for verifying the genuinely tripartite nonlocality of Werner state $\rho_{W}$.}
\label{fig-s7}
\end{figure}

{\bf Example S11}.  Consider a tripartite maximally entangled W state:  $|\Phi\rangle=\frac{1}{\sqrt{3}}
(|001\rangle+|010\rangle+|100\rangle)$. Define $A_i=\sigma_x$ for all $i$s. It is easy to check $\langle{}A_i\rangle_{\rho_{W}}=0$, and  $\langle{}A_i\otimes{}A_j\rangle_{\rho_{W}}=\frac{2v_3}{3}$ for all $i$s and $j\not=i$. From Eqs.(E51) and (E55) we get that
 \begin{align*}
\omega_{2\mbox{\small-pr}}(\rho_{W})=&\sum_{i\not=j}
(\frac{1}{3}
-\frac{4g_v}{9v_3}
\sqrt{\frac{1-v_3}{8}})\langle{}A_iA_j\rangle_{\rho_{W}}-\frac{2}{9}\frac{g_v^2}{v_3^2}(
\sum_{i=1}^3\langle{}A_i\rangle_{\rho_{W}})^2
\\
&
-\frac{4g_v}{9v_n}
\sqrt{\frac{1-v_n}{8}}\langle{}A_i\rangle_{\rho_{W}}
-\frac{2(1-v_n)}{3}
\\
=&\frac{4v_3}{3}-\frac{16g_v}{9}\sqrt{\frac{1-v_3}{8}}
-\frac{2(1-v_3)}{3}
\\
> &1
\tag{E57}
\end{align*}
which violates the inequality (E55) for $k=2$ when $v_3> v^*_3$, where $g_v=\sqrt{v_3+\frac{1-v_3}{8}}
-\sqrt{\frac{1-v_3}{8}}$. Numeric evaluations of $v$ are shown in Figure S\ref{fig-s7}.

\end{document}